\documentclass[fleqn,12pt]{elsarticle}

\usepackage[
colorlinks=true, pdfstartview=FitV, linkcolor=red, citecolor=blue, urlcolor=blue]{hyperref}

\usepackage{graphicx,color}
\usepackage{amsmath,amssymb}
\usepackage{slashbox}
\usepackage{feynmf}
\newcommand{\Slash}[1]{{\ooalign{\hfil/\hfil\crcr$#1$}}}
\newcommand{\tr}{{\rm tr}}

\newcommand{\Nc}{N_{\rm c}}

\newcommand{\lqcd}{\Lambda_{\rm QCD}}
\newcommand{\vp}{\vec{p}}
\newcommand{\vq}{\vec{q}}
\newcommand{\vk}{\vec{k}}

\newcommand{\vr}{\vec{r}}
\newcommand{\vP}{\vec{P}}

\newcommand{\vR}{\vec{R}}
\newcommand{\la}{\langle}
\newcommand{\ra}{\rangle}

\newcommand{\calL}{\mathcal{L}}
\newcommand{\calR}{\mathcal{R}}
\newcommand{\calH}{\mathcal{H}}

\newcommand{\calA}{\mathcal{A}}

\newcommand{\calD}{\mathcal{D}}
\newcommand{\calN}{\mathcal{N}}
\newcommand{\calM}{\mathcal{M}}
\newcommand{\calF}{\mathcal{F}}
\newcommand{\calK}{\mathcal{K}}

\newcommand{\calP}{\mathcal{P}}
\newcommand{\calW}{\mathcal{W}}

\newcommand{\rmd}{\mathrm{d}}
\newcommand{\rmi}{\mathrm{i}}
\newcommand{\rme}{\mathrm{e}}

\newcommand{\up}{\uparrow}
\newcommand{\down}{\downarrow}

\newcommand{\tq}{ \tilde{q} }

\newcommand{\luv}{\Lambda_{\scriptscriptstyle \rm UV}}
\allowdisplaybreaks

\usepackage[normalem]{ulem}  

\begin{document}
%
%
\begin{frontmatter}
\tnotetext[]{RBRC-1161, RIKEN-QHP-213}
\title{Mesons in strong magnetic fields: \\(I) General analyses}
\author[rbrc,riken] {Koichi Hattori}
\author[ccnu,uiuc] {Toru Kojo}
\author[fra] {Nan Su}
\address[rbrc] {RIKEN BNL Research Center, Brookhaven National Laboratory, Upton NY 11973, USA}
\address[riken] {Theoretical Research Division, Nishina Center, RIKEN, Wako, Saitama 351-0198, Japan}
\address[ccnu] {Key Laboratory of Quark and Lepton Physics (MOE) and Institute of Particle Physics, Central China Normal University, Wuhan 430079, China}
\address[uiuc] {Department of Physics, University of Illinois at Urbana-Champaign, 1110 W. Green Street, Urbana, Illinois 61801, USA}
\address[fra] {Institut f\"ur Theoretische Physik, Goethe-Universit\"at Frankfurt, 60438 Frankfurt am Main, Germany}

%
%
\begin{abstract}
We study properties of neutral and charged mesons in strong magnetic fields $|eB| \gg \lqcd^2$ 
with $\lqcd$ being the QCD renormalization scale. Assuming long-range interactions, 
we examine magnetic-field dependences of various quantities 
such as the constituent quark mass, chiral condensate, meson spectra, and meson wavefunctions 
by analyzing the Schwinger-Dyson and Bethe-Salpeter equations. 
Based on the density of states obtained from these analyses, 
we extend the hadron resonance gas (HRG) model to investigate thermodynamics at large $B$. 
As $B$ increases the meson energy behaves as a slowly growing function of the meson's transverse momenta, 
and thus a large number of meson states is accommodated in the low energy domain; 
the density of states at low temperature is proportional to $B^2$. 
This extended transverse phase space in the infrared regime
significantly enhances the HRG pressure at finite temperature, 
so that the system reaches the percolation or chiral restoration regime 
at lower temperature compared to the case without a magnetic field; 
this simple picture would offer a gauge invariant and intuitive explanation of the inverse magnetic catalysis.
\end{abstract}
\end{frontmatter}

\section{Introduction}
\label{sec:intro}

In uniform magnetic fields, a trajectory of a charged particle wraps around a magnetic flux, 
leading to the discretized orbital levels known as the Landau levels. 
Each level has degenerate states and the density of states is given by $ |eB|/2\pi$. 
Also, spin-$\frac{1}{2}$ fermions are subject to the Zeeman energy splitting, 
with which the $n$-th Landau level ($n$LL) has the following spectrum at the tree level,
\begin{equation}
E_n(p_3) = \sqrt{ p_3^2 + 2n|eB| + m^2} 
\label{eq:energy}
\, ,
\end{equation}
where the magnetic fields are applied to the $x_3$-direction, 
and $e$ and $m$ are the electric coupling constant and the fermion mass, respectively. 
A remarkable aspect for the fermionic case is that the energy of the lowest Landau level (LLL) at $n=0$ 
does not depend on the value of $B$, since the energy 
gained by the Zeeman effect cancels out the zero point energy arising from the orbital motion. Therefore {\it the $B$-dependence of the LLL energy appears only through the radiative corrections}. Those radiative corrections vary for different theories: our objective is to use such sensitivity to explore the structure of theories, especially Quantum Chromodynamics in the infrared (IR) regime. The characteristic scale is given by the QCD renormalization scale parameter, $\lqcd \sim 0.2\,{\rm GeV}$. (For an extensive review on magnetized systems, see Ref.~\cite{Miransky:2015ava}.)

The magnetic field we shall consider is strong, $|eB| \gg \lqcd^2$, perhaps much stronger than can be realized in experimental 
laboratories\footnote{Beyond laboratory experiments, larger magnetic fields may be achieved. At the stage of electroweak (EW) transition in the early universe, $|eB|$ could reach $\sim 1\,{\rm GeV}^2$ \cite{Vachaspati:1991nm}, but this size is much smaller than other scales relevant at early epoch. Magnetars have surface magnetic fields of electron mass scale and might have even larger magnetic fields at their cores by squeezing the surface fluxes, but how magnetic fluxes are distributed in magnetars remains an unsettled issue.}; the magnetic fields at RHIC and LHC are $|eB| \sim 0.04\,{\rm GeV}^2$ and $\sim 0.3\,{\rm GeV}^2$, respectively, and persist only for a short time \cite{Kharzeev:2007jp,Skokov:2009qp}. 
Thus the applicability of our arguments to phenomenology is subtle, but it is not our primary purpose. The strong field regime has been studied in the lattice simulations  \cite{Buividovich:2008wf,Bali:2011qj,Bornyakov:2013eya,Bali:2013esa,Bonati:2014ksa}, and we use them to test various theoretical concepts and methods of calculations, aiming at applications to other unexplored domain, e.g., physics of cold and dense QCD. 
Indeed, the similarities between the low energy dynamics of 
QCD at high density and in strong magnetic fields 
have been pointed out \cite{Gusynin:1994xp,Fukushima:2011jc,Kojo:2012js}. 
For example, the ``QCD Kondo effect'', that is induced by the infrared instability near the Fermi surface, 
was recently studied first in dense QCD \cite{Hattori:2015hka}, 
and then it was shown that the similarity between the Fermi surface effect and 
the (1+1) dimensional low energy dynamics in strong magnetic fields 
manifests itself as the ``magnetically induced QCD Kondo effect'' \cite{Ozaki:2015sya}.

Specific problems address in the present work are 
theoretical paradoxes observed in the recent lattice QCD studies. 
Two problems can be thought of as the representatives: (i) {\it The $B$-dependence of the chiral condensate} \cite{Buividovich:2008wf}: The chiral effective models or chiral perturbation theory confirmed the magnetic catalysis \cite{Shovkovy:2012zn} and explain the lattice data at $|eB| \ll \lqcd^2$ well, while beyond $|eB| \sim 0.1-0.3\,{\rm GeV}^2$, their predictions start to deviate from the lattice results which show the linear rising behavior, $\la \bar{\psi} \psi \ra \sim |eB| \lqcd$. 
(ii) {\it Inverse magnetic catalysis} \cite{Bali:2011qj,Bornyakov:2013eya}: The (2+1)-flavor lattice results 
at the physical pion mass show that the chiral restoration temperature ($T_c$) decreases at larger $B$, by $10-20\%$ at $|eB| \sim 1\, {\rm GeV}^2$, 
and this decreasing behavior continues to the maximal magnetic fields achieved on the lattice, $|eB| \simeq 3.3\, {\rm GeV^2}$ \cite{Endrodi:2015oba}. 
Typical model calculations \cite{Gatto:2010pt} instead predict the qualitatively opposite behavior: 
$T_c$ grows rather rapidly as $B$ increases. 

Effective models leading to these paradoxes have certain common features. 
Those models predict the dynamically generated quark mass, 
which is large ($\gg \lqcd$) and strongly $B$-dependent, 
e.g., $M \sim |eB|^{1/2}$ or $\sim \Lambda_{ {\rm UV} }\, \rme^{-c/|B| }$ 
($\Lambda_{ {\rm UV} }$ is the UV cutoff of the model). 
Let us see how this estimate causes the aforementioned problems. 
At large $B$, the proper fermion bases are the Ritus bases by using which 
one can include the background $B$ effects in the full order. 
The chiral condensate can be written as [$\int_{p_L} \equiv \int \rmd^2 p_L/(2\pi)^2$, $p_L=(p_0,p_3)$]
\begin{equation}
\la \bar{\psi} \psi \ra_{ {\rm 4D} } = \frac{\, |eB| \,}{2\pi} \la \bar{\psi} \psi \ra_{ {\rm 2D} } \,,
~~~~~  \la \bar{\psi} \psi \ra_{ {\rm 2D} } \equiv  - \int_{p_L} \tr\left[ S^{ {\rm 2D} }_{ {\rm LLL} } (p_L) + \sum_{n=1} S^{ {\rm 2D} }_{ n{\rm LL} } (p_L) \right] \,,
\end{equation}
where $S^{ {\rm 2D} }_{nLL}$ is the two-dimensional propagator for the $n$-th LL. Therefore if the mass gap rapidly grows as $B$ increases, the two-dimensional condensate $\la \bar{\psi} \psi\ra_{{\rm 2D} }$ also does. Then  $\la \bar{\psi} \psi \ra_{ {\rm 4D} }$ no longer depends on $B$ linearly, contradicting with the lattice results. The estimate of $M \gg \lqcd$ also causes a problem for the critical temperature, because with such large mass gap the thermal fluctuations of quarks are not activated until the temperature reaches $T \gg \lqcd$. Then the chiral restoration temperature keeps growing as $B$ increases, contradicting with the inverse magnetic catalysis. 
Due to the suppression of the quark fluctuations, 
it is also difficult to explain the $B$-dependence of various gluonic quantities 
observed in the lattice data \cite{Bali:2013esa,Bonati:2014ksa}. 
Note also that in the lattice simulations the inverse magnetic catalysis can be observed 
only when the quark mass is sufficiently small near the physical point \cite{Bali:2011qj}.

To cure these problems, several authors have emphasized fluctuation effects of various kinds, including hadronic fluctuations \cite{Fukushima:2012kc,Chao:2013qpa,Feng:2014bpa} or the back-reaction from quark to gluon sector \cite{Bruckmann:2013oba,Ferrer:2014qka,Farias:2014eca,Ferreira:2014kpa,Mueller:2015fka}. Several elaborated model studies took into account these fluctuations, and it became clear that models predicting $M \gg \lqcd$ do not reproduce the inverse magnetic catalysis \cite{Skokov:2011ib,Fukushima:2012xw,Kamikado:2013pya}. 
After all, the large fluctuation effects, that is expected to be necessary 
for the inverse magnetic catalysis phenomenon, are not well activated with such a large mass gap. 
Note also that the fluctuation arguments alone do not explain why the chiral condensate depends on $B$ linearly.

Keeping these problems in mind, we proposed that the quark mass gap can stay around $\sim \lqcd$ even for $|eB| \gg \lqcd^2$ \cite{Kojo:2012js}.
If this is true, we have $\la \bar{\psi} \psi \ra_{ {\rm 2D} } \sim \lqcd$ so that $\la \bar{\psi} \psi \ra_{ {\rm 4D} } \sim |eB|\lqcd$. In addition, at larger $B$ quark fluctuations are no longer suppressed, but enhanced due to the IR phase space extended by the magnetic fields. Indeed, a perturbative effective potential at a {\it fixed} effective quark mass indicates the reduction of $T_c$ as $B$ increases \cite{Ozaki:2013sfa}. All these features seem consistent with the lattice results. 

Of course, the question is how to get the mass gap of $\sim \lqcd$. The estimate strongly depends on the nature of the interactions. As argued in the previous works \cite{Kojo:2012js}, the $B$-dependence drops off from the mass gap if the IR contributions of the interactions dominate over the UV contributions. This observation is consistent with the Schwinger-Dyson studies in which the IR enhanced force was examined \cite{Kojo:2012js,Watson:2013ghq}. 
This feature is presumably very specific to non-Abelian gauge theories, not present in typical effective models.

One of the weakness in our arguments was that the quark self-energies are generally gauge-dependent quantities, and sometimes our intuition does not work well for such objects. For example, the quark self-energies can be IR divergent in Coulomb gauge type confining models \cite{'tHooft:1974hx}. These artifacts disappear at the level of color-singlet objects for which artificial contributions cancel each other, leaving only physical quantities. Thus in this paper we will try to deduce (or define) the constituent quark mass out of the hadron spectra; in fact this is how the concept of the constituent quark mass has been developed in traditional arguments. 

\begin{figure*}[!t]
\begin{center}
\includegraphics[width = 0.6\textwidth]{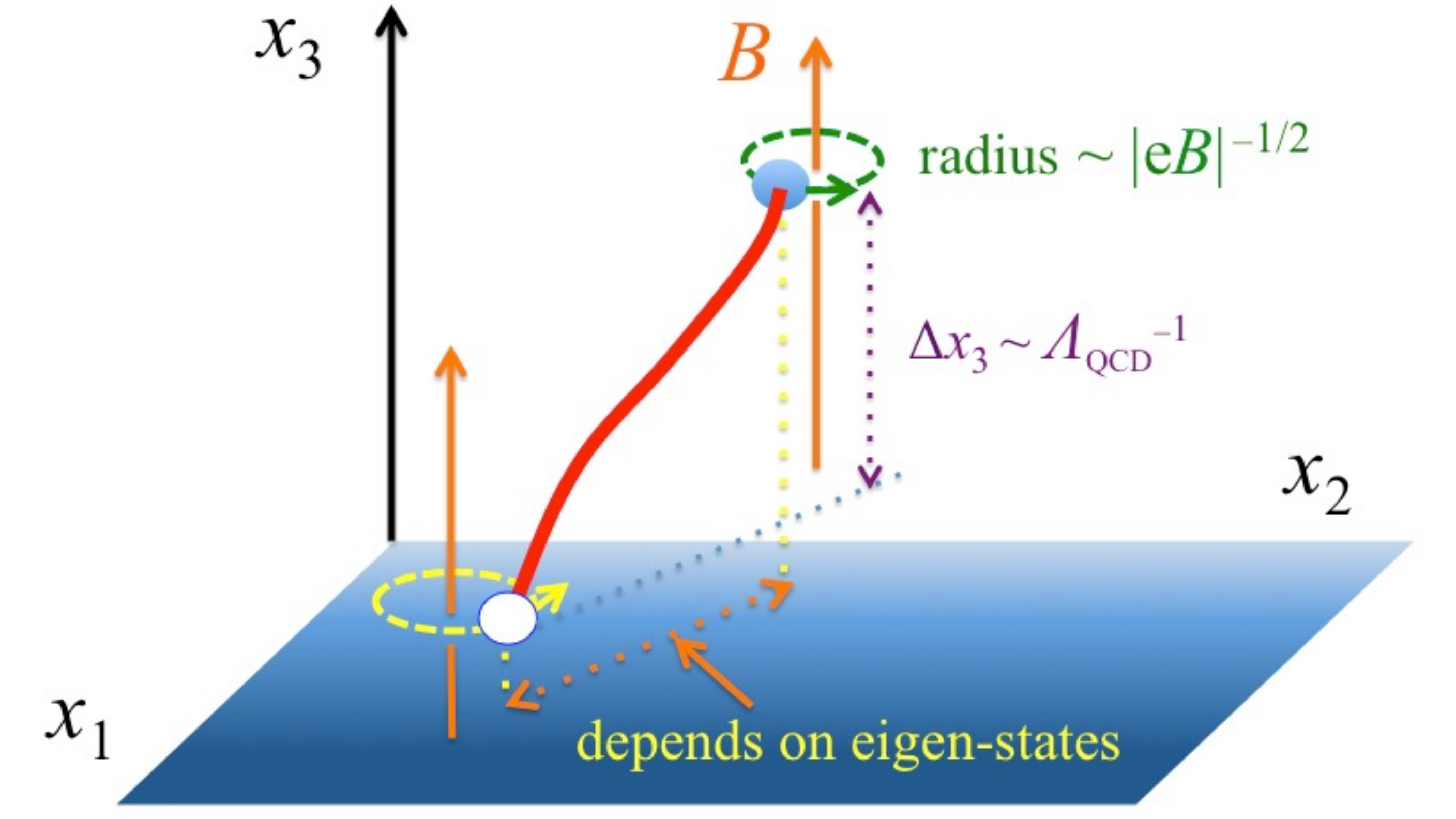}
\end{center}
\vspace{0.cm}
\caption{
\footnotesize{A schematic picture for a meson state to be discussed in this paper. A quark and an anti-quark wrap around magnetic fluxes and the radius of the trajectories is $\sim |B|^{-1/2}$. The distance between these two trajectories is $\sim \lqcd^{-1}$ in the direction of magnetic fields, while the transverse distance depends on the eigen-states; 
in low energy bound states the transverse distance are much smaller than $\lqcd^{-1}$, 
saving the length of the color-flux-tube.}
\vspace{-0.0cm}
}\label{fig:meson_intro}
\end{figure*}

This observation motivates us to study mesons in strong magnetic fields\footnote{In this paper
the QED part is essentially quenched, and the magnetic field is always uniform. The possibility of non-uniform 
distributions of vortices due to the $\rho$-meson condensates 
(see e.g. Ref.\cite{Chernodub:2011gs}) is omitted from the very beginning by our setup.}. 
In the present context, a sketchy argument was already presented in Ref.\cite{Kojo:2012js}, 
while in this paper we will give a more elaborate account of meson properties. 
We focus on mesons of light flavors with quantum numbers 
allowing a quark and an anti-quark to stay at the LLL. 
Other mesons that inevitably contain higher LLs at any time slice have the energy of $\sim \sqrt{|eB|}$ and will not be discussed. In this selection, neutral mesons as well as charged mesons come into our game, and they are as light as $\sim \lqcd$. Studies of neutral mesons require treatments beyond those by purely hadronic models, since at large $B$ magnetic fluxes penetrate hadrons and couple to quarks inside. This causes the structural change in neutral mesons \cite{Kojo:2012js,Fukushima:2012kc,Simonov:2012if,Alford:2013jva,Taya:2014nha,Gub}. For the lattice studies, see Refs.\cite{Hidaka:2012mz,Luschevskaya:2014lga}. As an example, in Fig.\ref{fig:meson_intro} we show a schematic picture of a meson state to be discussed in this paper (see Sec.\ref{sec:BSeq} for the outline).

We will study mesons at finite transverse momenta for construction of the hadron resonance gas (HRG) model at finite $B$. The weak field regime was studied in the purely hadronic description in Refs.~\cite{Endrodi:2013cs,Agasian:2008tb}, while our concern in this paper is the strong field regime which requires considerations at the quark level. Our approach has some overlap in philosophy with those in Refs. \cite{Fukushima:2012kc,Orlovsky:2013aya}, although the details are different.
We will find that the transverse momentum dependence in the meson spectrum 
is sensitive to the nature of the interactions. 
In particular, if the IR component dominates, the energy cost associated with the transverse momenta is suppressed by $\sim B^{-2}$ for neutral mesons and $\sim |B|^{-1}$ for charged mesons, and thereby the meson energy grows very slowly as a function of the transverse momentum. This allows many meson states to stay at low energy. At finite temperature, the HRG pressure grows with an increasing $B$, approaching the percolation or chiral restoration regimes at lower $T$ compared to the $B=0$ case. This would offer a {\it gauge invariant} description for the inverse magnetic catalysis.

For all these purposes, we analyze the structures of the Schwinger-Dyson and Bethe-Salpeter equations. Most of parametric estimates will be given assuming the long-range interactions, such as linear rising or harmonic oscillator potentials. The use of them provides illuminating predictions dramatically different from models of short-range interactions. 
We will address why such difference emerges.

Our analyses for the quark self-energy and bound state spectra are essentially based on the large $\Nc$ approximation in which the back reaction from the quark sector to the gluon sector can be ignored. For example, the magnetic field induced quark screening effects for long-range forces, such as the reduction of the string tension observed in the lattice QCD \cite{Bonati:2014ksa}, will not be manifestly taken into account. But those effects enhance the inverse magnetic catalysis behavior and are in line with our original proposal. In this paper we will show that even without these back reaction effects one can address a plausible mechanism to explain the inverse magnetic catalysis.

This paper is intended for the first part of a series of papers. In this paper we try to address generic aspects of meson problems, leaving more quantitative and model-dependent estimates for the second paper.

This paper is structured as follows. In Sec.~\ref{sec:preparation}, we introduce the Ritus bases for fermions, and discuss the $B$-dependent form factor and Schwinger phase which arise from the quark-gluon vertices. We argue why the LLL and higher LLs tend to decouple at large $B$ in asymptotic free theories. In Sec.~\ref{sec:SDeq}, we analyze general structures of the Schwinger-Dyson equation. In Sec.~\ref{sec:BSeq} we proceed to analyses on the structure of the Bethe-Salpeter equations for neutral and charged mesons. More specific aspects are elaborated in 
Secs.~\ref{sec:neutral} and \ref{sec:charged} for neutral and charged mesons, respectively. 
In Sec.\ref{sec:HRG}, we discuss the meson resonance gas, using the knowledges deduced in the previous sections. Sec.\ref{sec:summary} is devoted to summary and outlook.
In \ref{sec:app}, we discuss some operator relations which are useful to classify the meson states made of quarks in the LLL. 

\section{
QCD interactions in magnetic fields
}
\label{sec:preparation}

\subsection{Decomposition by Ritus bases}
\label{sec:Ritus}

In the presence of a magnetic field $B$, our tree level Langrangian reads 
\begin{equation}
\calL_{0} = \sum_f \bar{\psi}_f \left(\Slash{\calD} + m_f \right) \psi_f \,,
~~~~~~
\calD_\mu = \partial_\mu + \rmi e_f \calA_\mu \,,
\end{equation}
where $\psi_f$ is a quark field with the flavor index $f$ and the current quark mass $m_f$, 
and $\calA_\mu$ is a $U_{\rm em} (1)$ external gauge field. 
We use the Euclidean signature $g_{\mu \nu} = \delta_{\mu \nu}$ 
so that we do not have to distinguish $x_\mu$ and $x^\mu$. 
The $\gamma$-matrices are defined as Hermitian matrices, 
$\gamma_\mu^\dag = \gamma_\mu$ and 
$\gamma_5 = \gamma_0 \gamma_1\gamma_2\gamma_3=\gamma_5^\dag$.
Below we use notations, $x_L=(x_0,x_3)$, $x_\perp =(x_1,x_2)$, and 
the counterparts for momenta, 
assuming that a uniform magnetic field is applied in the $x_3$-direction. 
We choose the Landau gauge $\calA_\perp =(0, B x_1)$,
and will find that this choice simplifies an issue of gauge dependence 
in the Schwinger-Dyson equation. 
For later convenience, we also introduce the short-hand notations, 
$B_f \equiv e_f B$ and $s_f \equiv {\rm sgn}(B_f)$, 
where ${\rm sgn}$ is the sign function.

There are three steps to construct the Ritus bases. 
(i) The first step is to project out the spin of charged fermions 
as the energy levels in a magnetic field are subject to the Zeeman effect. 
This can be done by using the spin-projection operators 
$\calP_\pm^{f} = (1 \pm s_{f} \rmi \gamma_1 \gamma_2 ) / 2 $ as 
\begin{equation}
\psi^f_{\pm} \equiv \calP^{f}_\pm \psi^f \, ,
~~~ \rmi \gamma_1 \gamma_2 \, \psi_\pm^f 
= \pm \, s_{f}  \psi^f_\pm \, .
\end{equation}
It is important to remember that $\calP_\pm$ commutes with $\gamma_L$, 
while anti-commute with $\gamma_\perp$ which flips the spin.
(ii) Secondly, noting that momenta $p_L$ and $p_2$ are conserved in the Landau gauge, 
we can expand fermion fields as
\begin{equation}
\psi^f_\pm (x ) 
= \sum_{l=0}^\infty \int \frac{\rmd^2 p_L \rmd p_2 }{(2\pi)^3} \,
\psi^f_{\pm} ( l, p_2, p_L ) \,
\calH^f_l \!\left( x_1 - r^f_{p_2} \right) 
\,  \rme^{\rmi p_2 x_2} \, \rme^{\rmi p_L x_L } 
\label{eq:expand}
\,,
~~~~ r_{p_2}^f \equiv - \frac{p_2}{\, B_f \,}
\,,
\end{equation}
where the index $l$ labels an orbital level in the transverse motion. 
$\calH_l^f$ is the normalized harmonic oscillator function. 
In this work, we will use the explicit form at $l=0$ which is given by 
\begin{equation}
\calH_0^f (x_1-r^f_{p_{2} } ) 
= \left( \!\frac{\,  |B_f| \,}{\pi} \!\right)^{\!1/4}
\, \exp\left[ - \frac{\, |B_f| \,}{2} 
\left(x_1 - r_{p_2}^f \right)^2
\right]
\,.
\end{equation}
Note that the principal quantum number $n$ appearing in the energy level (\ref{eq:energy}) 
is the sum of the orbital level index $l$ and the Zeeman splitting $\pm g /2$
with the $g$-factor ($g=2$), i.e., $n = 2 l + 1 \pm 1$; 
the energy levels are two-fold degenerated except for the unique ground state $n=0$. 
(iii) Therefore, the last step is to relabel the fermion fields 
since the orbital level indices are not conserved numbers even at the tree level. 
We set 
\begin{equation}
\chi^f_{p_2} (p_L) \equiv \psi^f_{n=0, p_2}(p_L)
\equiv \psi^f_+(l=0, p_2, p_L) \,, 
\end{equation}
for the LLL, and
\begin{equation}
\psi^{f+}_{n, p_2} (p_L) \equiv \psi^f_+(l=n, p_2, p_L) \,, 
~~~
\psi^{f-}_{n, p_2} (p_L) \equiv \psi^f_- (l=n-1, p_2, p_L) \,,
\end{equation}
for the higher LLs (hLLs). 
We also define $\psi^f_{n, p_2} \equiv \psi^{f+}_{n,p_2} + \psi^{f-}_{n,p_2}$,
which will be useful to have a concise expression of the action below.

Then the zero-th order quark action for a flavor $f$ can be written 
as a sum of the Landau levels as $\calL^f = \calL^f_\chi + \sum_{n=1} \calL^f_n$,
where
\begin{equation}
\int_x \calL_\chi^f
=  \int_{p_L, p_2}
\bar{\chi}^f_{p_2} (p_L) \left( \rmi \Slash{p}_L + m_f \right) \calP^f_+ \chi^f_{p_2} (p_L)
\,, 
\end{equation}
for the LLL, and
\begin{equation}
\sum_{n=1}  \int_x \calL^f_n
=\sum_{n=1} \int_{p_L, p_2}
\bar{\psi}^f_{n,p_2} (p_L) \left( \rmi \Slash{P}^f_n + m_f \right) \psi^f_{n,p_2}
(p_L)
\,,
\end{equation}
for the hLLs with $(P_n^f)_\perp \equiv (\,0, s_f \sqrt{2n|B_f|} \,)$. 
Note that the $B$-dependence drops off from the tree-level Lagrangian for the LLL, 
as emphasized in Introduction.

\subsection{Quark-gluon vertices with form factors and Schwinger phases}
\label{formfactors}

While the Ritus bases diagonalize the tree-level Lagrangian, 
interactions induces transitions between different LLs at the quark-gluon vertices. 
In particular the $B$-dependence of the LLL appears only through the interactions. 
Using the expansion (\ref{eq:expand}) by the Ritus basis, 
the vertex of a $f$-flavored quark reads 
\begin{align}
\rmi g_s \int_x \,  \bar{\psi}_f \gamma_\mu t_a \psi_f  A_{\mu }^a (x)
&= \rmi \sum_{l,l'=0}^\infty
\int_{p_L,p_2,k_L,k_2} 
\bar{\psi}_f (l, p_2, p_L ) \, \gamma_\mu t_a\,
\psi_f (l', k_2 , k_L  ) \,
\nonumber \\
& \hspace{1cm}
\times
\int_{k_1} g_s A_{\mu}^a (p-k)
\int_{x_1} \calH^f_l(x_1-r^f_{p_2}) \calH^f_{l'} (x_1-r^f_{k_2}) \, \rme^{\rmi (p_1-k_1) x_1} 
\label{eq:vertex}
\,,
\end{align}
where $t_a$ is color matrices normalized as $\tr[ t_a t_b ] = \delta_{ab}/2$, 
and $A_\mu^a$ is a gluon field accompanying the coupling constant $g_s$.

We find in Eq.~(\ref{eq:vertex}) that the integral is factorized 
because the Ritus bases do not manifestly depend on $k_1$ and $x_1$. 
The last integral factor works as an overlap function 
among the plane-wave basis for a gluon and the Ritus bases for two quarks. 
Performing the integral, this factor can be written in a product form 
\begin{equation}
\int_{x_1} \calH^f_l(x_1-r^f_{p_2}) \calH_{l'} (x_1-r^f_{k_2}) \, \rme^{\rmi (p_1-k_1) x_1} 
\equiv
\rme^{ \rmi \Xi^f_{p,k} } \, \calR^f_{ll'}(\vp_\perp - \vk_\perp)\,.
\end{equation}
The phase factor $\Xi^f_{p,k}$ is called the Schwinger phase, 
and its explicit form is given by 
\begin{equation}
\Xi^f_{p,k} \equiv (p_1-k_1) r^f_{(p_2+k_2)/2} = - (p_1-k_1) \frac{\, p_2 + k_2 \,}{2B_f } 
\label{eq:Sch}
\, ,
\end{equation}
which depends upon $(p_2+k_2)/2$, that is, 
the average of fermion momenta before and after the interaction. 
This exponent is gauge variant; the residual symmetry in the Landau gauge, $\calA_2=Bx_1 \rightarrow B (x_1+c)$ preserves magnetic fields but shifts the fermion momenta in the $x_2$-direction. 
An advantage in the Landau gauge is that 
the exponent is common to 
all of the combinations of couplings between different LLs\footnote{
This simplification is by no means trivial. In fact, in the symmetric gauge, it is difficult to disentangle the Schwinger's phase part and the others; the expansion of the Schwinger's phase appears as powers of the pseudo-angular momenta, and apparently the fermions 
labeled by different pseudo-angular momenta acquire different self-energies. 
The Schwinger's phase can be factored out only after summing up all powers of 
the pseudo-angular momenta that yields the exponentiated form.}. 
We will see that the Schwinger phase provides physical effects when we calculate many-body states.

The other factor at the vertex depends only on the momentum transfer flowing into the gluon line as 
\begin{equation}
\calR^f_{ll'} (\vp_\perp - \vk_\perp)
\equiv
\int_{x_1} \calH^f_l \left( x_1-r^f_{(p_2-k_2)/2} \right) \, \calH^f_{l'} \left(x_1 + r^f_{ (p_2-k_2)/2} \right) \, 
\rme^{\rmi (p_1-k_1) x_1} \,.
\end{equation}
We call it the form factor. For the momentum transfer $\vq_\perp=\vp_\perp - \vk_\perp$, 
we find 
\begin{equation}
\calR_{ll'} (\vq_\perp) \propto 
\left\{ \begin{matrix}
& \left( \frac{ q_1 - \rmi s_f q_2}{ | 2B_f |^{1/2} }\right)^{l-l'} &~~~( l>l')\\
& \left( \frac{ q_1 +\rmi s_f q_2}{ | 2B_f |^{1/2} }\right)^{l'-l} &~~~( l'>l)
\end{matrix}\right. \,.
\end{equation}
When $l\neq l'$, these power factors weaken the relevance of interactions 
in the IR regime, $q_{1,2} \lesssim \sqrt{B_f}$. 
Thus the $l\neq l'$ processes, in which a quark hops from $l$-th to $l'$-th orbital level, 
are dominated by the UV interactions. In asymptotic free theories, 
these UV interactions are in the weak-coupling regime 
and can be treated within the perturbation theory. 
On the other hand, such IR suppressions do not occur for the $l=l'$ processes, 
so that results are sensitive to the IR structure of the interactions. 
Therefore, one must directly handle the non-perturbative features of gluon propagators 
and quark-gluon vertices explicitly.

\subsection{Interactions for the LLL}

\begin{figure*}[!t]
\begin{center}
\includegraphics[width = 0.4\textwidth]{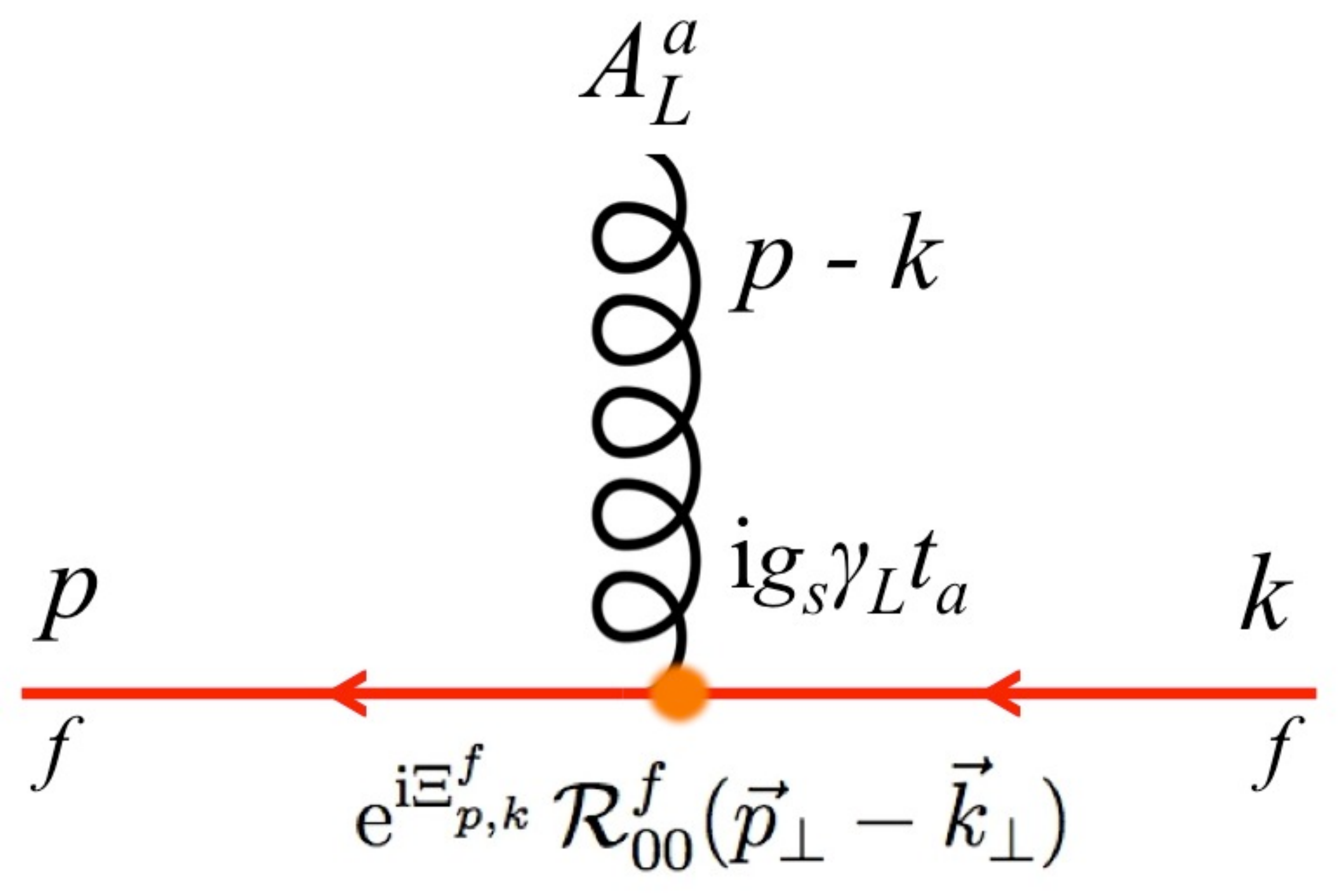} 
\end{center}
\vspace{-0.5cm}
\caption{
\footnotesize{The quark-gluon vertex for the LLL with the flavor $f$.
}
\vspace{-0.4cm}
}\label{fig:vertex}
\end{figure*}

In this work we focus on the LLL in the strong field limit. 
As depicted in Fig.~\ref{fig:vertex}, 
the interaction Lagrangian is given by
\begin{equation}
\calL_{ {\rm LLL} }^{ {\rm int} }
= \rmi \sum_f
\int_{p_L,p_2,k_L,k_2} 
\bar{\chi}^f_{p_2} ( p_L ) \, \gamma_L t_a\, \calP^f_+ \chi^f_{k_2} ( k_L  ) \,
\int_{k_1}  \rme^{ \rmi \Xi^f_{p,k} } \, \calR^f_{00}(\vp_\perp - \vk_\perp)
\times g_s A_{L}^a (p-k) \,.
\end{equation}
Note that the $\gamma_\perp$ component drops off because $\calP_+ \gamma_\perp \calP_+ =0$. 
More physically, the $\gamma_\perp$ vertices flip spins and inevitably couple the LLL to hLLs, 
so that it is irrelevant when we consider only the low energy sector. 
This observation also suggests that the relevant gauge fields at the low energy are $A_L$, 
while $A_\perp$'s tend to decouple from the low energy dynamics. 
The lattice simulations for gluon condensates and string tensions 
have indeed shown that configurations of $A_L$ are more screened than $A_\perp$, 
and this behavior can be understood from the suppression of 
the coupling to $A_\perp$ at the LLL \cite{Bali:2013esa,Bonati:2014ksa}.

A final comment is on the form factor. One finds its explicit form at the LLL as 
\begin{equation}
R_{00}^f ( \vq_\perp) = \rme^{ - \frac{ \vq_\perp^2 }{\, 4|B_f| \,} } 
\label{eq:ff}
\,,
\end{equation}
which cuts off the UV contribution when $\vq_\perp^{\, 2}$ is larger than $|B_f|$. 
In other words, the size of $|B_f|$ determines to what extent the UV sector 
can participate in the LLL dynamics. There is another important remark: 
at $\vq_\perp^{\, 2} \ll |B_f|$, the form factor behaves as
\begin{equation}
R_{00}^f ( \vq_\perp) \simeq 1 - \frac{ \vq_\perp^{\,2} }{\, 4|B_f| \,} + \cdots \,,~~~~(\vq_\perp^{\,2} \ll |B_f|)
\, ,
\end{equation}
which means that the $B$-dependence can be dropped off from the IR sector. 

With these observations, now we can see that there are two distinct contributions in the interactions, 
namely, the magnetic-field {\it dependent} contributions from the UV interactions, 
and magnetic-field {\it independent} contributions from the IR interactions. 
The $B$-dependence of the quark self-energies, chiral condensates, meson spectra, and so on, is determined by the competition between the UV and IR contributions, 
and thus is very sensitive to the fundamental nature of theories one considers. 
For instance, if one employs the models involving contact interactions, 
the UV contributions are strongly enhanced, 
and then the quark self-energies are very sensitive to the value of $B$. 
In contrast, theories involving stronger IR interactions show a much weaker $B$-dependence. 
Therefore, in this respect, QCD should be distinguished from the models 
that do not properly include the dynamics of the gluonic sector.

\section{Schwinger-Dyson equations for the LLL: General analyses}
\label{sec:SDeq}

\begin{figure*}[!t]
\begin{center}
\includegraphics[width = 0.7\textwidth]{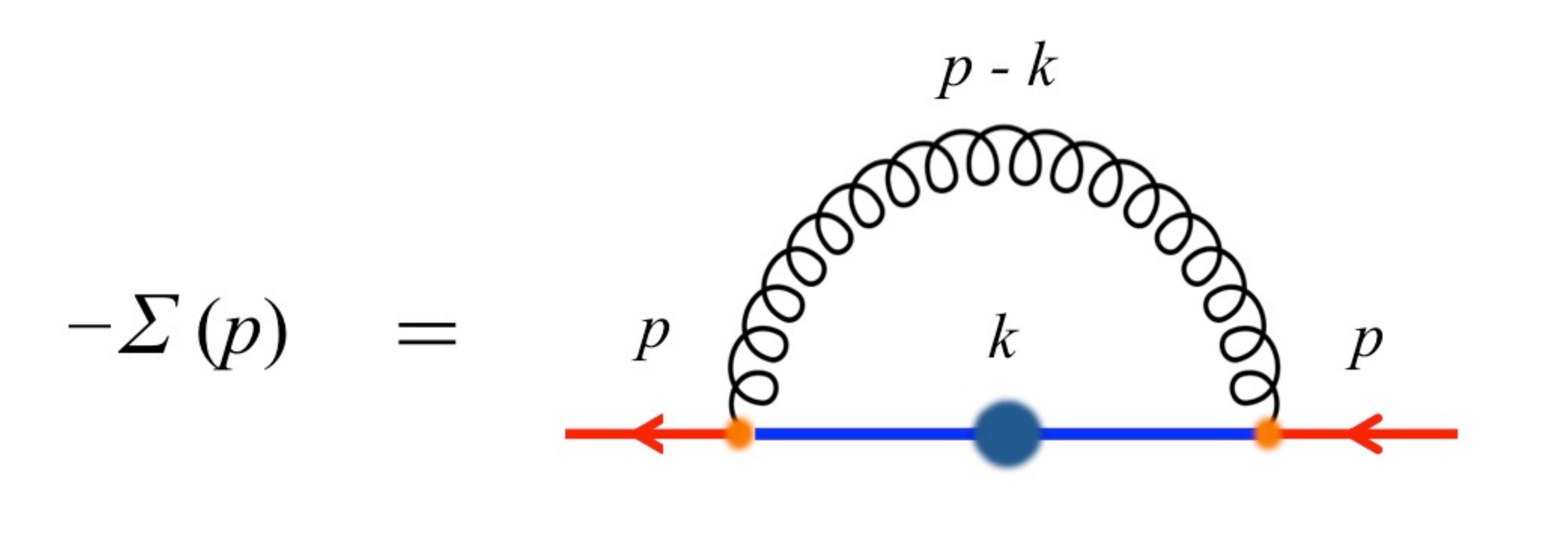} 
\end{center}
\vspace{-0.3cm}
\caption{
\footnotesize{The Schwinger-Dyson equation within the rainbow approximation. 
}
\vspace{-0.4cm}
}\label{fig:SDeq}
\end{figure*}

This section is devoted to general analyses on the Schwinger-Dyson equations 
for the self-energy of the LLL. 
For simplicity we consider the rainbow approximation only as shown in Fig. 3, 
but we expect that our arguments hold even after this 
restriction is removed. 
Assuming the strong field limit, 
we restrict both the external and intermediate quark states to the LLL. 
For $|eB| > 0.3\,{\rm GeV}^2$, the impact of hLL intermediate states 
to the LLL self-energies was found to be small\footnote{
The large energy gap between the LLL and the hLL itself does not justify the LLL approximation, 
because an infinite sum of small contributions of the order of $ |B|^{-1}$ produces a UV divergence. 
However, this UV divergence 
should be nothing but the divergence in the vacuum at $B=0$. 
Therefore, the divergence can be absorbed into the renormalized current quark mass at $B=0$, 
and the remaining hLL contributions result in only (very) small higher order $B$-corrections \cite{Kojo:2013uua}. 
In short, the summation of discretized energy levels and the integral of continuum states 
yield similar results, so details of the hLL contributions do not play a major role in the LLL self-energies.}, 
so we may ignore it \cite{Kojo:2013uua}.

The dressed quark propagator for the LLL can be written as 
\begin{equation}
S(q_L, {\bf \Sigma} (q) ) 
= \left[\, \rmi (\Slash{q} +\Slash{\Sigma} )_L + (m+\Sigma_m) \,\right]^{-1} \calP_+
= S_0 + S_0 (-{\bf \Sigma} ) S_0 + \cdots
\,,
\end{equation}
where $S_0 (q) = (\rmi \Slash{q}_L + m)^{-1}$ is the tree propagator of the LLL, 
and for the moment we suppress flavor indices. 
The self-energy $\Sigma$ is decomposed into 
\begin{eqnarray}
{\bf \Sigma} \equiv \rmi \Slash{\Sigma}_L + \Sigma_m
\, .
\end{eqnarray}
We have dropped off the $\Sigma_\perp$ component 
because the $\gamma_\perp$ structure is forbidden 
by the projection operators as $\calP_+ \gamma_\perp \calP_+ =0$. 
Denoting the gluon propagator as $D^{ab}_{\mu \nu} = \delta^{ab} D_{\mu \nu}$, 
the Schwinger-Dyson equation is then given by
\begin{equation}
- {\bf \Sigma}(p)
= (-\rmi)^2 C_F \int_{k_L,\vk_\perp} \gamma_{L'} S(k_L; {\bf \Sigma} (k)) \gamma_{L''}
\times D_{L' L''} (p-k) \left[ \calR_{00}(\vp_\perp - \vk_\perp) \right]^2 
\label{eq:SD}
\,,
\end{equation}
where $C_F=(\Nc^2-1)/2\Nc$ is the Casimir for the fundamental representation. 
We will keep the general form of the gluon propagator $D_{\mu \nu}$ 
which includes insertions of gluon self-energies and is convoluted with dressed vertices.
We emphasize that in the self-energy calculations the Schwinger phases 
from two vertices exactly cancel out.

Next, we will show that the self-energy is independent of the transverse momenta even after radiative corrections are included. To prove it, we compare the self-energies $\Sigma(p_L,\vp_\perp)$ and $\Sigma(p_L, \vp_\perp+\delta \vp_\perp)$. First we notice that, in contrast to the $B=0$ cases, 
the propagator depends on $p_\perp$ only through the self-energy part. 
Secondly, the integrand other than the propagator part depends only on the difference $\vp-\vk$. 
Thus, by shifting the integration variable $\vk_\perp$, we find 
\begin{align}
&\hspace{-0.3cm}
{\bf \Sigma}(p_L,\vp_\perp+\delta \vp_\perp) - {\bf \Sigma}(p_L,\vp_\perp)
\nonumber \\
&\hspace{-0.3cm}
\propto
\int_{k_L,\vk_\perp}
\left[\, S[k_L; {\bf \Sigma} (k_L,\vk_\perp+\delta \vp_\perp)] 
- S[k_L; {\bf \Sigma} (k_L,\vk_\perp) ] 
\, \right] \times 
D_{L' L''} (p-k) \left[ \calR_{00}(\vp_\perp - \vk_\perp) \right]^2 
\,.
\end{align}
This relation must hold for any values of $\delta \vp_\perp$. 
An obvious candidate is $\Sigma(p_L, \vp_\perp) = \Sigma(p_L)$, 
which is independent of $\vp_\perp$; in this case the LHS and RHS both vanish in the above equation. 
This is a natural solution as the perturbative calculation also results 
in the $\vp_\perp$-independent self-energy. 

With the above solution, 
we may write ${\bf \Sigma}(q) = {\bf \Sigma} (q_L)$, 
and then find that the integral equation (\ref{eq:SD}) is factorized as follows. 
The separation between 
the longitudinal and transverse momentum dependences 
allows us to define a two-dimensional effective propagator by 
\begin{equation}
D^{ {\rm 2D} }_{L' L''} (p_L-k_L) 
\equiv C_F
\int_{\vk_\perp} D_{L' L''} (p-k) 
\left[ \calR_{00}(\vp_\perp - \vk_\perp) \right]^2 \,,
\end{equation}
and the dimensionally reduced Schwinger-Dyson equation 
in the (1+1) dimensions by 
\begin{equation}
\rmi \Slash{\Sigma}_L (p_L) + \Sigma_m (p_L) 
=  \int_{k_L} \gamma_{L'} S(k_L) \gamma_{L''} \times D^{ {\rm 2D} }_{L' L''} (p_L-k_L) \,.
\end{equation}
(We should remember that the flavor indices for ${\bf \Sigma}$, $S$, and $D^{ {\rm 2D} }$ are suppressed.)

In this equation, the only one origin of the $B$-dependence 
is the form factor hidden in the definition of $D^{ {\rm 2D} }$, 
since the LLL propagator does not contain any manifest $B$-dependence. 
Therefore our problem to examine the $B$-dependence is now reduced 
to the study of the effective propagator; 
if the effective propagator strongly depends on $B$, the self-energies also do; 
if not, we will find the self-energies (nearly) independent of $B$, 
which, as mentioned in Introduction, provides the key ingredient of the scenario 
for the magnetic catalysis and inverse catalysis of the chiral condensate 
observed by the lattice QCD simulations.

\section{The Bethe-Salpeter equations for the LLL: General analyses}
\label{sec:BSeq} 

\begin{figure*}[!t]
\begin{center}
\includegraphics[width = 0.7\textwidth]{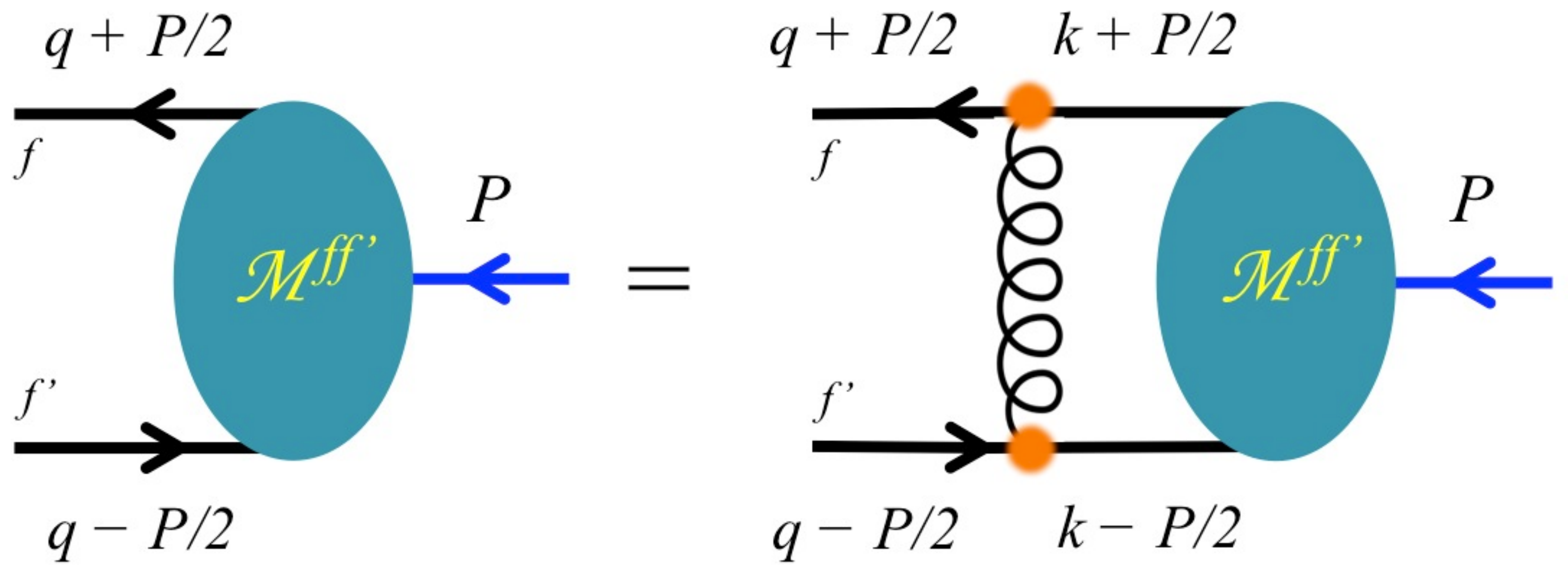} 
\end{center}
\vspace{-0.3cm}
\caption{
\footnotesize{The (homogeneous) Bethe-Salpeter equation within the rainbow ladder approximation.
}
\vspace{-0.4cm}
}\label{fig:BSamp}
\end{figure*}

\subsection{Some definitions}

In this section we investigate general structures of the Bethe-Salpeter equations 
for bound states composed of the fermions in the LLL, and 
discuss characteristic differences between neutral and charged mesons. 
The 
self-energies discussed in the previous section 
is supposed to be used as an input to the Bethe-Salpeter equation.
We consider the Bethe-Salpeter amplitude defined by 
\begin{equation}
\la 0 |\,\bar{\chi}^{f' }_{q_2^-} (q_L^-) \chi^{f }_{q_2^+} (q_L^+) \, | \calM_{P'_2 P'_L} \ra 
\equiv
(2\pi)^3 \, \delta(P_2-P_2')\, \delta(P_3-P_3')\, \delta\left(\rmi P_0+E(P_3') \right)
\times \calM^{f f' }_{P_2 P_3}  ( q_2, q_L )  \,,
\end{equation}
where $(P_2,P_L)$ are conserved total momenta, 
and $q_L^\pm = q_L \pm P_L/2$ and $q_2^\pm = q_2 \pm P_2/2$ 
are momenta carried by a quark and an anti-quark, respectively (see Fig.\ref{fig:BSamp}). 
Especially we will be interested in the equal-time amplitude, which is defined as
\begin{equation}
\calM^{f f' }_{P_2 P_3}  ( q_2, q_3 )  \equiv  \int_{q_0} \calM^{f f' }_{P_2 P_3}  ( q_2, q_L ) \,,  
\end{equation}
in which the $\chi^{f'}$ and $\chi^f$ hit the meson state at the same time. 
Using relative and center-of-mass (COM) coordinates, 
$r_i = x_i - x'_i$ and $R_i = (x_i +x_i')/2$, respectively, 
the coordinate-space expression can be constructed as 
\begin{equation}
\calM^{ff'}_{P_2 P_3} (\vr; \vR) \equiv \calN \, \rme^{ \rmi ( P_2 R_2 + P_3R_3) } \int_{q_2, q_3} 
\rme^{ \rmi (q_2 r_2 + q_3 r_3) } \calH_0^{f} \left(x_1 - r^f_{ q_2^{+} } \right) \calH_0^{f'} \left(x_1'-r^{f'}_{q_2^-} \right)  
\calM^{f f' }_{P_2 P_3}  ( q_2, q_3 ) 
\label{eq:M}
\,,
\end{equation}
where $\calN$ is some normalization factor. 
(Here we included $\rme^{\rmi P_2 R_2}$ in the definition since the integral still contains $P_2$ which will be combined with the exponential factor.)

From this form of the wavefunction, we can already get some insights on the bound state problem. 
At a large $B$, the probability distributions in the $x_1$-direction are localized around $r^f_{q_+}$ and $r^{f'}_{q_-}$ with the variances of $\sim 1/|B|^{1/2}$, reflecting the cyclotron motion of each quark.  Therefore the relative and COM coordinates are roughly given by combinations of the conserved momentum $P_2$ and the expectation value of the relative momentum $\la q_2 \ra$ as
\begin{align}
& \la R_1 \ra \sim \left( \frac{1}{\, B_f \,} + \frac{1}{\, B_{f'} \,}  \right) \la q_2 \ra + \left( \frac{1}{\, B_f \,} - \frac{1}{\, B_{f'} \,}  \right) \frac{\, P_2 \,}{2} \,,
\label{eq:R1}
\\
& \la\, r_1 \, \ra  \sim \left( \frac{1}{\, B_f \,} - \frac{1}{\, B_{f'} \,}   \right) \la q_2 \ra + \left( \frac{1}{\, B_f \,} + \frac{1}{\, B_{f'} \,}  \right) \frac{\, P_2 \,}{2} 
\label{eq:r1}
\,.
\end{align}
To get intuitive pictures for our bound state problem, first we stress that the momenta we are dealing with
are the ``canonical'' momenta which are gauge-variant.
After making a gauge choice, this momentum is related to the ``guiding center'' around which particles exhibit the cyclotron motion.
More intuitively appealing quantity is the ``kinetic'' momentum, $p + e_f A$, which is related to the velocity.
For instance a particle at large $x_1$ has large canonical momentum $k_2 \simeq - B_f x_1$ 
but also large $e_f A_2= B_f x_1$. 
As the sum of them, the kinetic momentum is not large 
and remains $\sim |B_f |^{1/2}$ as expected from the cyclotron motion; 
a particle with large $k_2$ has neither large velocity nor large Lorentz force. 

Except the quantum aspects related to the vector potential and guiding center, the remaining part can be understood within the classical picture.
Like the (classical) Hall effect,
a particle (anti-particle) drifts in the direction orthogonal to those of magnetic fields and a force acting on it.
The direction of the drift depends on the electric charge. 
Now let us examine two characteristic examples.

\begin{figure}[!t]
\begin{center}
\includegraphics[width = 0.8\textwidth]{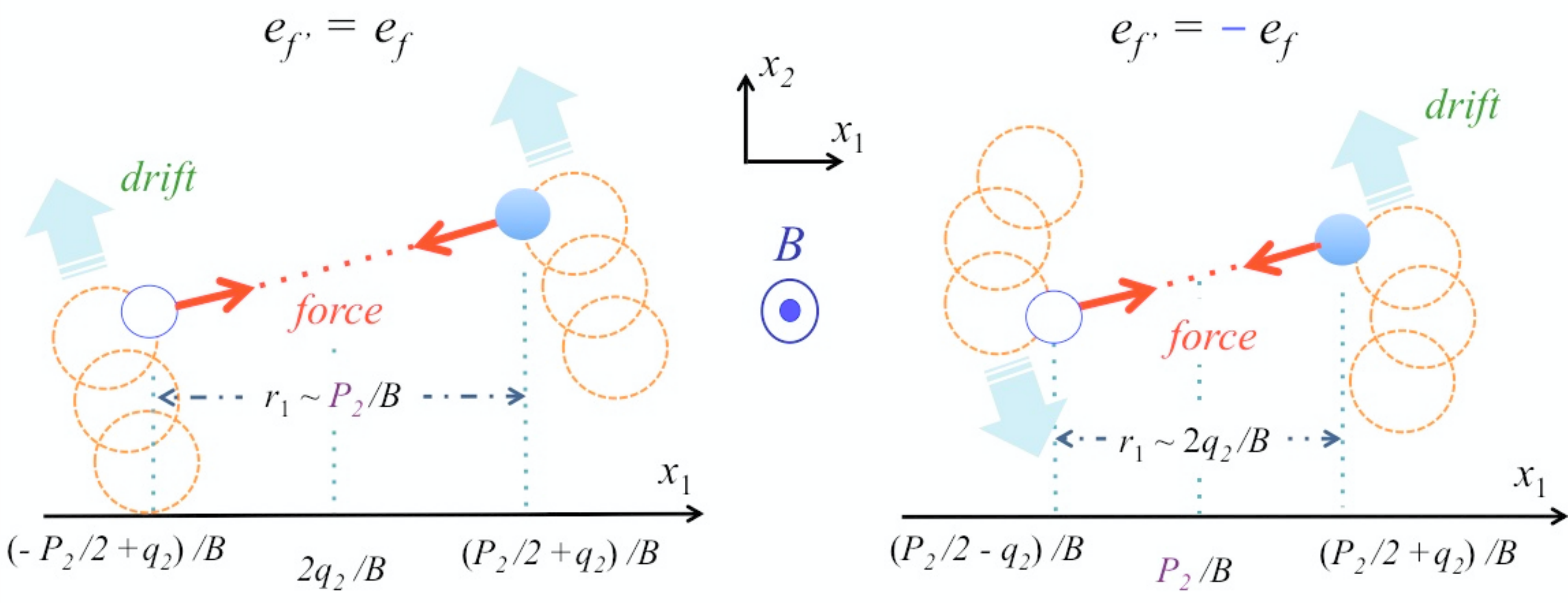} 
\end{center}
\vspace{-0.5cm}
\caption{
\footnotesize{A semi-classical picture for mesons projected onto the plane perpendicular to magnetic fields. (Left) The $e_{f'} =e_f$ case for neutral mesons. A particle and an anti-particle attract each other and drift in the same direction orthogonal to those of the force and magnetic fields (Hall drift). The constants of motion are the relative coordinates with which the strength of the two-body interaction is kept fixed.
(Right) The $e_{f'} = - e_f$ case for charged mesons. The directions of the drift are opposite for a particle and an anti-particle, changing the relative coordinate. The constant of motion is the COM in the $x_1$-direction ($P_2/B$), and the typical mode with such conserved quantity is the rotation.
}
\vspace{-0.cm}
}\label{fig:Hall_drift}
\end{figure}

In case of $e_f = e_{f'}$ for the neutral mesons [Fig.\ref{fig:Hall_drift}(left)], we have 
\begin{equation}
\la R_1 \ra_{e_f= e_{f'} } \sim \frac{\, 2 \la q_2 \ra \,}{B_{f} } \,,~~~~~~ \la r_1 \ra_{ e_f = e_{f'} } \sim \frac{\, P_2 \,}{ B_{f} } 
\label{eq:Rr_N}
\,.
\end{equation}
In this case, the relative momentum is related to the COM coordinate.  
We find that the neutral meson spectra are degenerated with respect to this momentum 
since the spectra are independent of the location of the COM coordinate in a constant magnetic field. 
On the other hand, the conserved total momentum $P_2$ manifestly enters the bound state problem 
because, as seen in Eq.~(\ref{eq:Rr_N}) and Fig.\ref{fig:Hall_drift}, it corresponds to the relative distance 
between a particle and an anti-particle.
In the classical picture, the relative distance is naturally kept fixed since the particle and anti-particle
drift in the same direction with the same velocity, as a consequence of $e_f=e_{f'}$.
Moreover this also indicates that besides the conserved momentum $P_2$, there should be
another constant of motion that is related to the relative distance in the $x_2$-direction.
Indeed such conserved quantity appears in our Bethe-Salpeter equation as shown below. 

Another characteristic case is bound states of equal charges $e_f = -e_{f'}$ (Fig.\ref{fig:Hall_drift}(right)), 
which does not exist in the mesonic system (because of 
the difference between $e_u =2e/3$ and $e_d = -e/3$) 
but can be applied to diquark or dielectron systems. 
In such cases, $q_2^- = q_2 -P_2/2$ should be replaced by $\tq_2^- = -q_2 +P_2/2$, 
and this procedure is equivalent to taking $e_{f'}=-e_f$ in the transverse dynamics of 
the mesonic systems. Then, from Eqs.~(\ref{eq:R1}) and (\ref{eq:r1}), we find 
\begin{equation}
\la R_1 \ra_{e_f= -e_{f'}} \sim \frac{\, P_2 \,}{B_{f} } \,,~~~~~~ \la r_1 \ra_{e_f= - e_{f'}} \sim \frac{\, 2 \la q_2 \ra\,}{ B_{f} } \,.
\end{equation}
The spectra again should not depend on the location of $R_1$, 
so that the spectra are degenerated with respect to $P_2$ 
in contrast to the case of neutral mesons discussed above. 
On the other hand, the relative distance $\la r_1 \ra$ depends on 
the non-conserved momentum $q_2$, leading to nontrivial oscillating modes 
which couple the motions in the $x_1$- and $x_2$-directions.
In the classical picture, a particle and anti-particle drift in the 
opposite direction while keeping their COM coordinate in the $x_1$-direction at $P_2/B_f$;
this motion around a fixed point can be interpreted as a rotating mode 
whose radius becomes large for higher excitations. 

Below we will provide more precise statements by analyzing the mathematical structure of the Bethe-Salpeter equation, and will see that the above heuristic arguments indeed capture the essence of the bound state problems.

\subsection{The Bethe-Salpeter equations in the Rainbow Ladder approximation}

In the Rainbow Ladder approximation, 
the Bethe-Salpeter equation for a pair of fermions in the LLL states is written down as 
\begin{align}
&\hspace{-0.cm}
\calM^{ff'}_{P_{2}P_3 }  (q_2, q_L) 
= (-\rmi)^2 C_F \, 
\nonumber \\
&\hspace{0.5 cm}
 \times 
S^f (q_L^+ ) \, \gamma_L \left[
 \int_{k_L, k_\perp} \!\! D_{LL'} (q- k) \,
\calF^{ff'}_{q_2^+,k_2^+ ; \, q_2^-,k_2^-} (q_1-k_1) \,
\calM^{ff'}_{P_{2} P_3 } (k_2,k_L) \right] 
\gamma_{L'}  S^{f'} (q_L^-) 
\label{eq:BS}
\,,
\end{align}
where the factor $\calF$ summarizes the Schwinger phases (\ref{eq:Sch}) 
and the form factors (\ref{eq:ff}) as 
\begin{equation}
\calF^{ff'}_{q_2^+,k_2^+ ; \, q_2^-,k_2^-} (q_1-k_1) 
\equiv
\calR_{00}^f  \calR_{00}^{f'} (\vq_\perp-\vk_\perp) \times
\rme^{ \rmi \left[ \Xi^f_{q_2^+,k_2^+} (q_1-k_1) \, - \, \Xi^{f'}_{q_2^-,k_2^-} (q_1-k_1) \right] } \,.
\end{equation}
%
Before examining roles of $\calF$, 
let us first note that the $B$-dependence enters the LLL dynamics only through the factor $\calF$. 
Further we find that the exponents of form factors and Schwinger phases appear 
with an inverse factor of $1/\vert B \vert$ 
as $ (\vq_\perp -\vk_\perp )^2/|B| $ and $(q_1-k_1)/|B|$, respectively. 
Therefore, if one plays with theories of the enhanced IR interactions, 
that factor can be set to $\calF \simeq 1$, 
and then the spectra are of the order of $ \lqcd$ independently of $B$.

The above situation may, however, alter for meson states with large momenta. 
In such cases, the Schwinger phases play a dramatically important role. 
For mesons composed of quarks with $f$ and $f'$ flavors, 
the quantum phase takes the form
\begin{equation}
 \left[ \Xi^f_{q_2^+,k_2^+} - \Xi^{f'}_{q_2^-,k_2^-} \right] (q_1-k_1) 
= - (q_1-k_1) \times \left\{ 
\frac{q_2+k_2}{2} \left[ \frac{1}{B_f} - \frac{1}{B_{f'}} \right]
+  \frac{P_2}{2} \left[ \frac{1}{B_f} + \frac{1}{B_{f'} } \right]
\right\} 
\label{eq:phase}
\,.
\end{equation}
Note that the above expression depends on both the total momentum $P_2$ and $(q_2+k_2)/2$ 
which is the average relative momentum before and after the interaction. 
Clearly, this factor manifestly couples the COM dynamics to the relative dynamics between two particles, i.e., 
the COM and relative motions are not independent of each other. 
In the subsequent sections, we will examine how this factor plays a role  
in the $e_f = e_{f'}$ and $e_f  \neq e_{f'}$ cases more specifically.

\section{Neutral mesons : the $e_f = e_{f'}$ case}
\label{sec:neutral}

For $e_f=e_{f'}$, the quantum phase (\ref{eq:phase}) in the last section becomes
\begin{equation}
 \left[ \Xi^f_{q_2^+,k_2^+} - \Xi^{f}_{q_2^-,k_2^-} \right] (q_1-k_1) 
= - (q_1-k_1) \frac{P_2}{ B_f } 
\label{eq:phase_N}
\,,
\end{equation}
where the dependence on the average relative momentum $(q_2+k_2)/2$ drops off, 
while the total momentum $P_2$ remains. 
There is another special character in the $e_f=e_{f'}$ case. 
As shown below, we can define a (fictitious) transverse momentum $P_1$, 
in addition to the conversed momentum $P_2$ which already exists in Eq.~(\ref{eq:phase_N}).

By making use of the additional momentum $P_1$, 
we can transform the Bethe-Salpeter equation into a rotationally symmetric form. 
To see this, we shall first determine the $q_2$-dependence of the amplitude. 
To simplify expressions, we suppress irrelevant indices and arguments such as $q_L, k_1,\cdots$, 
and write the structure of the Bethe-Salpeter equation in the following symbolic form 
\begin{equation}
\calM (q_2) = \int_{k_2} \calK (q_2- k_2) \calM (k_2) \,.
\end{equation}
%
The kernel $\calK$ depends on $q_2$ and $k_2$ only through the difference $q_2-k_2$. 
This is because (i) the Schwinger phase does not depend on $q_2+k_2$ specifically for 
the $e_f=e_{f'}$ case, and (ii) the fermion propagators are independent of $k_2$ and $q_2$. 
The latter is a special character of the Landau level propagators. 
With this symbolic form at hand, one can conclude that if $\calM(q_2)$ satisfies the equation, 
then $\calM(q_2+\delta u)$ also does, because
\begin{equation}
\calM (q_2+\delta u) = \int_{k_2} \calK (q_2 + \delta u - k_2) \calM (k_2) = \int_{k_2} \calK (q_2 - k_2) \calM (k_2 +\delta u) \,.
\end{equation}
Therefore $\calM(q_2)$ and $\calM(q_2+\delta u)$ 
characterize the same eigen-states. 
These degenerated states are labeled 
by the aforementioned momentum $P_1$ in the following way.
These two states are different from each other only by a constant phase, 
\begin{equation}
\calM(q_2+\delta u) = \rme^{ \rmi C_{\!\scriptscriptstyle \delta u} } \calM(q_2) \,.
\end{equation}
%
On the other hand, one can write
\begin{equation}
\calM ( q_2+\delta u) = \rme^{ \delta u \partial_{q_2} } \calM (q_2)\,,
\end{equation}
so that the solution is found to be 
\begin{equation}
\calM (q_2) = \rme^{ \rmi q_2 C_{\delta u}/\delta u} \, \calM (q_2=0) 
\label{eq:translation}
\,.
\end{equation}
The constant factor $C_{\delta u}/\delta u$ characterizes the amplitude and spectrum of the bound state. 
Now, using a fictitious transverse momentum $P_1 = - C_{\delta u}/\delta u$, 
we may label the states connected by the translation (\ref{eq:translation}) as 
\begin{equation}
\calM_{P_1} (q_2) \equiv \rme^{ -\rmi q_2 P_1/B_f } \, \calM_{P_1} (q_2=0) 
\label{eq:P1}
\, .
\end{equation}
%
This relation indicates that the shift of the relative momentum $q_2$ affects on 
neither the total momenta $P_1$ nor $P_2 $. 
Substituting the above expression for the previous symbolic form, one finds
\begin{equation}
\calM_{P_1} (q_2=0) = \int_{k_2} \calK (q_2- k_2) \, \rme^{ \rmi (q_2-k_2) P_1/B_f } \calM_{P_1} (k_2=0) \,.
\end{equation}
This new phase factor for $P_1$ can be combined with the factor from the Schwinger phase, yielding
\begin{equation}
-\rmi (q_1-k_1)\frac{P_2}{B_f} + \rmi (q_2-k_2)\frac{P_1}{B_f} 
= \rmi \, \frac{\vec{B_f} \times \vec{P}_\perp }{\, {B_f^2} \,} \cdot (\vq -\vk)_\perp 
\equiv \rmi \, \vec{\xi}_P^{\,f}  \cdot (\vq -\vk)_\perp  \,,
\end{equation}
which is rotationally symmetric with respect to $\vP_\perp$. 
Now recovering all the other indices in the original expression (\ref{eq:BS}), 
the Bethe-Salpeter equation can be written in the dimensionally reduced form
\begin{equation}
\calM^{f}_{P_{\perp }P_3 }  (q_L) 
= - S^f (q_L^+ ) \, \gamma_L \left[
 \int_{k_L} \!\! \calW^{f }_{LL'} (q_L- k_L; P_\perp) \,
\calM^{f}_{P_{\perp} P_3 } (k_L) \right] 
\gamma_{L'}  S^{f} (q_L^-) 
\,,
\end{equation}
where $\calM_{P_\perp P_3} (q_L) \equiv \calM_{P_\perp P_3} (q_2=0, q_L)  =\rme^{\rmi q_2 P_1/B_f} \calM_{P_\perp P_3} (q_2, q_L) $, and 
we have defined the (1+1)-dimensional force
\begin{equation}
\calW^{f}_{LL'} (q_L-k_L;P_\perp)
\equiv C_F \int_{k_\perp} D_{LL'} (q-k)\,
\left[ \calR_{00}^f  (\vq_\perp-\vk_\perp) \right]^2 \times
\rme^{ \rmi \, \vec{\xi}_P^{\,f} \cdot ( \vq_\perp-\vk_\perp )  } \,.
\end{equation}

To get some insights on the impact of the quantum phase, 
it is instructive to look at the expression in the coordinate space. 
With the inverse Fourier transform, the above expression becomes
\begin{equation}
\calW^{f}_{LL'} (r_L;P_\perp)
= C_F \, \frac{\, |B_f | \,} {2\pi} \int_{ \vr_\perp} D_{LL'} (r_L, \vr_\perp) \, \rme^{- \frac{ |B_f| }{2} \left(\vr_\perp - \vec{\xi}_P^{\, f} \right)^2 } \,.
\end{equation}
If the variation in $D_{LL'}$ is much milder than in 
$\rme^{- \frac{ |B_f| }{2} \left(\vr_\perp - \vec{\xi}_P^{\, f} \right)^2 }$, 
we can use a factorization to get
\begin{equation}
\calW^{f}_{LL'} (r_L;P_\perp)
\simeq  C_F \, D_{LL'} \left(r_L, \vr_\perp = \vec{\xi}_P^{\, f} \right) \,,
~~~~~~~~~~~ \frac{\, |B_f | \,} {2\pi} \int_{ \vr_\perp}  \rme^{- \frac{ |B_f| }{2} \left(\vr_\perp - \vec{\xi}_P^{\, f} \right)^2 } = 1\,,
\end{equation}
where we have replaced the $\vr_\perp$ in the potential by $\vec{\xi}_P$, and then carried out the integration. 
As an important example, 
we find that the linear rising potential results in 
\begin{eqnarray}
\calW^{f}_{00} (r_L;P_\perp) \sim \sigma \sqrt{ r_3^2 +  ( \vec{\xi}_P^{\, f} )^2  }
\, .
\label{eq:1D_n}
\end{eqnarray}  
For such an interaction, $\vec{\xi}_P^f$ can be dropped off in the limit of large $B \gg P_2$, 
and the problem is reduced to the bound state problem with a potential $\sim \sigma |r_3|$ 
in one spatial dimension; 
in this case the bound states are squeezed into 
the one-dimension shape and their spectra are independent of $B$. 

The above qualitative picture of the bound states will be considerably changed 
if the mutual interaction has short-range nature 
where our factorization approximation may not be valid. 
For a contrast example, a contact interaction gives the following one-dimensional potential, 
\begin{equation}
D_{ {\rm contact} } (\vr)  \sim - \lqcd^{-2}\, \delta( \vr)  ~~~ \rightarrow ~~~
\calW_{ {\rm contact} } (r_3 ; P_\perp)
\simeq - \frac{ |B_f| }{\, \lqcd^2 \,}\, \rme^{- \frac{P_\perp^2}{ 2 |B_f| } } \, \delta( r_3)
\,,
\end{equation}
which strongly depends on $B$ and the binding energy grows as $B$ increases. 
If one uses another interaction, e.g., $D \sim 1/\sqrt{r_3^2+\vr_\perp^2}$, 
such a strong $B$-dependence would be tempered. 
However, in this case, some special care is necessary for the $r_3^2 \sim 0$ region 
where the $r_\perp$-dependence and thereby $\xi_P$-dependence again become significant. 
For these cases, details of $\vec{\xi}_P^f$ become very important, 
and the spectra will be sensitive to the value of $B$. 
Having compared the bound states by three simple interactions, 
we find that the bound state spectra are strongly model-dependent, 
and especially depend on whether the interactions have short- or long-range nature.


Finally we shall examine the structure of the equal-time amplitude in the coordinate space. 
Following from Eqs.~(\ref{eq:M}) and (\ref{eq:P1}), we find the coordinate-space expression as 
\begin{equation}
\calM^{f}_{P_\perp P_3} (\vr; \vR) 
\equiv 
\calN\, \rme^{ \rmi (P_2 R_2 +P_3 R_3) } \int_{q_2} 
\rme^{ \rmi q_2 r_2 } \, \calH_0^{f} \left(x_1 - r^f_{q_2^+} \right) \calH_0^{f} \left(x_1'-r^f_{q_2^-} \right)  
\, \rme^{-\rmi q_2 P_1/B_f} \, \calM^{f}_{P_\perp P_3}  ( r_3 ) \,,
\end{equation}
where the inverse Fourier transform of $\calM(q_3)$ 
is denoted as a normalized one-dimensional wavefunction $\calM^f_{P_\perp P_3} (r_3)$. 
After carrying out the $q_2$-integration, we arrive at
\begin{equation}
\calM^{f}_{P_\perp P_3} (\vr; \vR)
= 
\frac{\, \rme^{\, \rmi \vP \cdot \vR \,} \,}{\, \sqrt{V_3} \,} \times
\, \rme^{-\rmi B_f R_1 r_2} \times
\sqrt{ \frac{\, |B_f| \,}{\, 2\pi \,} }\, 
\rme^{-\frac{|B_f|}{4} \left(\vr_\perp - \vec{\xi}^{\,f}_P \right)^2 } \,\calM^f_{P_\perp P_3} (r_3)\,,
\end{equation}
with $V_3$ being the three-dimensional volume. 
Several remarks are in order: (i) The first factor is the plane wave part characterized 
by conserved momenta, $\vP = (P_1,P_2,P_3)$. 
A factor of $\rme^{\rmi P_1 R_1}$ for $P_1$ was obtained after integrating over $q_2$. 
(ii) The Schwinger phase appears in the second factor because our interpolating fermion fields 
are separated in distance; if we started with a manifestly gauge invariant operator, $\bar{\chi} (x) \rme^{ \rmi e \int^x_y \rmd \vec{s} \cdot \vec{A} (s) }\chi (y)$, then the Schwinger phase could be eliminated. (iii) The third factor is the Gaussian function which characterizes 
the wavefunction for the relative transverse coordinate. 
The average transverse distance between the quark and anti-quark is given by $\vec{\xi}_P^f \sim B_f^{-1}$, 
which is vanishing as a magnitude of $B$ increases, 
again indicating the squeezed bound state.

\section{Charged mesons : the $e_f \neq e_{f'}$ case}
\label{sec:charged}

Next we turn to the $e_f \neq e_{f'}$ case. Since the COM and relative motions are correlated, 
the dynamics is more involved than the $e_f = e_{f'}$ case in the last section. 
First we find the quantum phase (\ref{eq:phase}) as 
\begin{align}
 \left[ \Xi^f_{q_2^+,k_2^+} - \Xi^{f'}_{q_2^-,k_2^-} \right] (q_1-k_1) 
= - 
a (q_1-k_1) 
\left[ \left( q_2 + \zeta P_2 \right) + \left( k_2 + \zeta P_2  \right) \right]  
\, ,
\end{align}
where we introduced 
%
\begin{eqnarray}
a = \frac{\, B_{f'} - B_f \,}{ 2 B_{f'} B_f } , 
~~~ \zeta = \frac{ 1}{2} \frac{ B_{f'} + B_f }{  B_{f'} - B_f  }
\, .
\end{eqnarray}
Note that, when the relative momentum is $q_2 -  \zeta P_2 $, 
the $P_2$-dependence in the exponent is absorbed by the Bethe-Salpeter amplitude, 
i.e., the Bethe-Salpeter equation (\ref{eq:BS}) can be rewritten as
\begin{align}
\calM_{P_2 P_3}^{ff'} \left( q_2 - \zeta P_2 , q_L\right)
=(-\rmi)^2 C_F 
S^f (q_L^+ ) \, \gamma_L 
 \int_{k_L, k_\perp}
& \!\! D_{LL'} (q- k) \,
 \calR_{00}^f  \calR_{00}^{f'} (\vq_\perp-\vk_\perp) \,
 \rme^{ - \rmi a (q_1 - k_1)(q_2 + k_2) } 
\nonumber 
\\
&
\times  
\calM_{P_2 P_3}^{ff'} \left( k_2 - \zeta P_2, k_L \right) 
\gamma_{L'}  S^{f'} (q_L^-) 
\,.
\end{align}
%
It is important to notice that the $P_2$-dependence appears 
only through the amplitude function $\calM$; the other part is $P_2$-independent, 
meaning that the amplitudes at $P_2=0$ and $P_2 \neq 0$ obey the same equation. 
Therefore the solutions at $P_2=0$ and $P_2 \neq 0$ 
are physically equivalent, and are equal modulo some constant phase factor 
(here we may regard $P_2$ as a constant because it is conserved): 
\begin{equation}
\calM_{P_2 P_3}^{ff'} \left( q_2 - \zeta P_2 , q_L\right) 
= \rme^{ \rmi C_{\!\scriptscriptstyle P_2} } \, \calM_{P_2=0, P_3}^{ff'} \left( q_2 , q_L\right) \,.
\end{equation}
We find that the COM and relative motions are correlated 
because the shift of $q_2$ affects on the $P_2$, and {\it vice versa}. 
This contrasts to the neutral case discussed around Eq.~(\ref{eq:P1}).

To get solutions for all $P_2$, we have only to analyze the Bethe-Salpeter equation at $P_2 =0$ 
and use the above relation. Therefore from now on we consider the $P_2=0$ case 
and shall drop off the subscript $P_2$. The remaining calculations are worked out as follows. 
Taking the coordinate-space expression of the propagator and 
then carrying out the integration over $k_1$, we arrive at
\begin{equation}
 \calM_{P_3}^{ff'} ( q_2, q_L )
= (-\rmi)^2 S^f (q_L^+ ) \, \gamma_L \, \left[
\int_{k_L} \int_{ r_L } \rme^{ - \rmi (q_L-k_L) r_L} \, 
\left( \calK \calM \right)_{P_3}^{ff'} \right] \gamma_{L'}  S^{f'} (q_L^-) \,,
\end{equation}
where $\calK$ assembles the transverse part,
\begin{equation}
\left( \calK \calM \right)_{P_3}^{ff'} 
=  \sqrt{ 4b\pi \,}\, C_F \int_{\vr_\perp} \int_{k_2} D_{LL'} (r_L, \vr_\perp)\, 
\rme^{ - b \left[\, r_1 + a (q_2+k_2) \,\right]^2 }
\rme^{-\rmi (q_2 -k_2) r_2 - \frac{1}{\, 4b \,} (q_2 - k_2)^2 }  \calM_{P_3}^{ff'} (k_2, k_L)
\label{eq:KM}
\, .
\end{equation}
In addition to $a  \sim B^{-1} $ introduced above, we have 
\begin{equation}
b = \frac{ |B_{f'} B_f| }{\, |B_{f'}| + |B_f| \,} = {\mathcal O} (B) \,.
\end{equation}

To proceed to further analytic computation, we focus on long-range interactions 
such as the linear rising potential, and make several approximations suitable 
for the analysis of those important models. 
(i) Below we assume that the propagator varies much more slowly 
than the Gaussian $\rme^{-b[\, r_1 +a(k_2+q_2) \,]^2}$, 
which tends to be the case in strong magnetic fields. 
Then we can apply a factorization, 
\begin{equation}
 \int_{r_1} D_{LL'} ( r_L, \vr_\perp ) \, \rme^{ - b \left[\, r_1 + a (q_2+k_2) \,\right]^2 }
\simeq D_{LL'} \! \left(\, r_L, r_1 = -a(q_2+k_2), r_2 \, \right) \times \sqrt{ \frac{\, \pi \,}{b} } \,,
\end{equation}
where $r_1$ is replaced by the center of the peak, $-a(q_2+k_2) \sim - (q_2 + k_2)/B$.
(ii) Assuming a long-range interaction, that is dominated 
by small momentum transfer processes $|q_2-k_2| \sim \lqcd \ll |B|^{1/2}$, 
we find 
\begin{equation}
\rme^{ - \frac{1}{\, 4b \,} (q_2 - k_2)^2 } \simeq 1 \,, 
~~~~~~~~~ a (q_2 + k_2) \sim 2a k_2 + O(\lqcd/B) \,,
\end{equation}
so that Eq.~(\ref{eq:KM}) reads 
\begin{equation}
\left( \calK \calM \right)_{P_3}^{ff'} 
 \simeq  2\pi\, C_F  \int_{r_2} \rme^{-\rmi q_2  r_2 } \int_{k_2}
D_{LL'} \left(\, r_L, r_1 = - 2a k_2, r_2 \, \right)\,
\rme^{ \rmi k_2 r_2 } \, \calM_{P_3}^{ff'} (k_2, k_L) \,.
\end{equation}
Now that the $q_2$-dependence was factorized in the exponential factor, 
our Bethe-Salpeter equation, in the coordinate representation with respect to $r_2$, is obtained as  
\begin{align}
 \calM_{P_3}^{ff'} ( r_2, q_L )
&\simeq (-\rmi)^2 S^f (q_L^+ ) \, \gamma_L \, 
\int_{k_L} \int_{ r_L } \rme^{ - \rmi (q_L-k_L) r_L} \, 2\pi C_F
\nonumber \\
& \times \left[ \int_{k_2} D_{LL'} \left(\, r_L, r_1 = - 2a k_2, r_2 \, \right)\,
\rme^{ \rmi k_2 r_2 } \, \calM_{P_3}^{ff'} (k_2, k_L) \right] \gamma_{L'}  S^{f'} (q_L^-) 
\, .
\label{BS_appro}
\end{align}
This approximate form is valid for long-range interaction models whose 
spatial variations in the coordinate $r_1$ is much milder than the Gaussian form factor
$\sim \rme^{-|B| r_1^2}$.

The expression (\ref{BS_appro}) has a somewhat involved structure, and requires some heuristic explanations. 
Because we are considering the low energy spectra in confining theories, 
the average $|\vr|$ should be ${\mathcal O} (\lqcd^{-1} )$. 
Especially, $\la r_3 \ra$ cannot be much smaller than $\lqcd^{-1}$ as in the case of $B=0$, 
because of the kinetic energy in the $x_3$-direction. 
However, as illustrated in Fig.~\ref{fig:meson_intro}, 
the transverse size, $\la r_1 \ra$ and $\la r_2\ra$, can be much smaller than $\lqcd^{-1}$. 
When one of the coordinates $\la r_1 \ra $, that is $\sim a \la k_2 \ra $ in Eq.~(\ref{BS_appro}), 
is much smaller than $\lqcd^{-1}$, 
the wavefunction should strongly damp beyond $k_2 \sim |B|/\lqcd$. 
On the other hand, when $|\la r_2 \ra|$ is small, 
the wavefunction should contain the Fourier components from zero to values much larger than $\lqcd$; 
otherwise, i.e., if only soft components $( \lesssim \lqcd)$ are included, 
we can get only $| \la r_2 \ra| > \lqcd^{-1}$. These two observations indicate 
that, for low energy states, the damping scale, $\Lambda_{ {\rm damp} }$, of the transverse wavefunction 
should satisfy a hierarchy 
\begin{equation}
\lqcd \ll \Lambda_{ {\rm damp} } \ll |B|/ \lqcd \,,~~~~~~~~~ \calM(k_2,k_L) \sim 0 ~~~~~( {\rm for}~ |k_2| \gg \Lambda_{ {\rm damp} })
\, .
\end{equation}
Furthermore, note that the relative distance 
in the transverse direction, $\la r_\perp \ra$, is minimized when
\begin{equation}
\la r_\perp^2 \ra \sim \la k_2^2 \ra /B^2 + \la r_2^2 \ra  \, \ge \, 2 \left( \la k_2^2 \ra \la r_2^2 \ra \right)^{1/2} \!/ |B|  
\label{eq:size}
\,,
\end{equation}
where the equality holds when $\la r_2^2 \ra = \la k_2^2\ra/B^2$. 
This relation is satisfied when $\la k_2^2 \ra \sim 1/\la r_2^2 \ra \sim |B|$, 
and such expectation values can be obtained for wavefunctions 
such as $\calM(k_2) \sim \rme^{ - k_2^2/|B|}$.
Such a solution with $\la r_1^2 \ra \sim \la r_2^2 \ra \sim |B|^{-1}$ 
is a natural candidate for the ground state because $\sim \la r_\perp^2 \ra$ is the minimal area
for the circular motion that was expected from the classical picture in Fig.\ref{fig:Hall_drift} (right).
At $|B| \gg \lqcd^2$, charged mesons at low energy are squeezed in a nearly one dimensional shape. 
For the $n_\perp$-th excited states, the transverse area grows as $\sim \la r_\perp^2 \ra_{n_\perp} \sim n_\perp |B|^{-1}$.
Therefore the transverse extension is negligible until $n_\perp$ reaches $\sim |B|/\lqcd^2$,
and there are many squeezed charged mesons at low energy. 

\subsection{Example 1: Confining harmonic oscillator}

We examine an instantaneous potential of the harmonic oscillator. 
Using this solvable model, we learn basic structure of the low energy bound state, 
and extend it to analysis of more general potentials. 
The explicit form of the potential is given by 
\begin{equation}
D^{ {\rm harmo} }_{LL'} (r) = \delta_{L0} \delta_{L'0} \, \delta(r_0) \, D^{ {\rm harmo} } (\vr)\,,
~~~~~D^{ {\rm harmo} } (\vr)= \frac{\, c_H \lqcd^3 }{2\pi C_F \,} \left(\, r_3^2 + r_\perp^2 \,\right) \,,
\end{equation}
where $c_H$  is some dimensionless~constant. In this model, 
we can rewrite the transverse part of 
the Bethe-Salpeter equation (\ref{BS_appro}) as
\begin{equation}
  \int_{k_2} \left(\, r_3^2 + \frac{k_2^2}{\, |B|^2 \,} + r_2^2 \,\right)  \rme^{\rmi k_2 r_2}\, \calM_{P_3}^{ff'} (k_2, k_L) 
=  \left(\, r_3^2 - \frac{ \partial_2^2}{\, |B|^2 \,} + r_2^2 \,\right) \calM_{P_3}^{ff'} (r_2, k_L) 
\,.
\end{equation}
The above operator can be diagonalized by using the harmonic oscillator bases; 
for the $n_\perp$-th excited state in the transverse directions, 
the corresponding wavefunction is found to be $\calM_{n_\perp} (r_2,r_L) = \calM_{n_\perp} (r_L) \, \calH_{n_\perp} (r_2)$, 
e.g., for the ground state, $\calM_0 (r_2,r_L) = \calM_0 (r_L) \, \rme^{-\frac{|B|}{2} r_2^2}$. 
Making use of these bases, the transverse dynamics is solved as 
\begin{equation}
  \left(\, r_3^2 - \frac{ \partial_2^2}{\, |B|^2 \,} + r_2^2 \,\right) \calM_{n_\perp} (r_2, k_L) 
  =  \calH_{n_\perp} (r_2) \left(\, r_3^2 + \frac{\, 2n_\perp+1 \,}{\, |B| \,} \,\right) \calM_{n_\perp} (k_L) \,.
\end{equation}
%
Inserting this expression into Eq.~(\ref{BS_appro}), 
the Bethe-Salpeter equation 
reads 
\begin{equation}
 \calM_{n_\perp, P_3}^{ff'} ( q_3 )
\simeq (-\rmi)^2 \int_{q_0} S^f (q_L^+ ) \, \gamma_0 \, 
\left[ \int_{k_3} D_{n}^{ {\rm harmo} } (q_3 - k_3)  \calM_{n_\perp, P_3}^{ff'} (k_3) \right] \gamma_{0}  S^{f'} (q_L^-) 
\, ,
\end{equation}
where, for the $n_\perp$-th excited transverse modes, 
\begin{equation}
D_{n_\perp}^{ {\rm harmo} } (q_3 ) = c_H \lqcd^3 \int_{r_3} \rme^{\rmi q_3 r_3}  \left(\, r_3^2 + \frac{\, 2n_\perp+1 \,}{\, |B| \,} \right) 
\label{eq:D_harmo}
\, .
\end{equation}
Since the instantaneous potential is independent of the temporal components, 
the $q_0$-integral results in the equal-time amplitude, $\calM(q_3) = \int_{q_0} \calM(q_L)$. 
In Eq.~(\ref{eq:D_harmo}), the second term, $(2n_\perp +1)/|B|$, is irrelevant 
until the integer $n_\perp$ reaches $\sim |B|/\lqcd^2$. 
Therefore, for $n_\perp \lesssim  |B|/\lqcd^2$, 
our problem is reduced to a one-dimensional bound state problem with respect to the relative coordinate $r_3$. 
This result indicates that there are a lot of (nearly) degenerated bound states at low energy 
whose energies are determined solely by the one-dimensional dynamics in the $x_3$-direction.

\subsection{Example 2: Confining linear rising potential}
Next we proceed to a more realistic confining potential, 
that is, an instantaneous linear rising potential 
\begin{equation}
D^{ {\rm linear} }_{LL'} (r) = \delta_{L0} \delta_{L'0} \, \delta(r_0) \, D^{ {\rm linear} } (\vr)\,,
~~~~~D^{ {\rm linear} } (\vr)= \frac{\, \sigma \,}{2\pi C_F \,} \sqrt{\, r_3^2 + r_\perp^2 \,} \,,
\end{equation}
with the string tension $\sigma \sim \lqcd^2$. 
We shall again analyze the transverse part in Eq.~(\ref{BS_appro}), 
\begin{equation}
  \int_{k_2} \sqrt{\, r_3^2 + \frac{k_2^2}{\, |B|^2 \,} + r_2^2 \, }  \,\, \rme^{\rmi k_2 r_2}\, \calM_{P_3}^{ff'} (k_2, k_L) \,.
\end{equation}
While it is tempting to replace $k_2$ by $-\rmi \partial_2$ as before, 
this is not a legitimate treatment since $\partial_2$ and $r_2$ inside the square-root do not commute. 
Thus the analysis is more involved than the harmonic oscillator case. 

Nevertheless, we may use an approximation and extend the procedure examined in the last subsection 
to analyze low energy states which satisfy the relation 
$|\la r_\perp \ra| \lesssim |B|^{-1/2} \ll  | \la r_3 \ra|  \sim \lqcd^{-1} $ 
discussed around Eq.~(\ref{eq:size}). 
We also note that the potential energy becomes important only when 
$| \la r_3 \ra| \gtrsim \lqcd^{-1}$, otherwise the potential energy term is simply negligible. 
In such a domain of $r_3$ and $r_\perp$, we can expand the potential as 
\begin{equation}
 \sim \int_{k_2} |r_3| \left[\, 1 + \frac{1}{\, 2|r_3^2 B| \,} \left( \frac{k_2^2}{\, |B| \,} + |B| r_2^2  \right) + O(1/ |r_3^2 B|^2 ) \, \right] \, \rme^{\rmi k_2 r_2}\, \calM_{P_3}^{ff'} (k_2, k_L) \,,
\end{equation}
where the expression is valid only for $|r_3| \gg |B|^{-1/2}$. 
In the second term, we can make replacement $k_2 \rightarrow -\rmi \partial_2$, 
and diagonalize it by using the harmonic oscillator bases. 
Corrections to these leading terms 
are of the order of $(\lqcd^2/|B|)^2  \ll 1$. 
Picking up a class of the higher order corrections, 
one can replace $\sqrt{r_3^2 + k_2^2/ |B| + r_2^2 \,}$ by 
$\sqrt{r_3^2 - \partial_2^2/|B| + r_2^2 \,} +O(\lqcd^4/|B|^2 )$. 
Within this precision level, one can expand $\calM(q_2)$ by the harmonic oscillator bases, and obtains 
\begin{equation}
 \calM_{n_\perp, P_3}^{ff'} ( q_3 )
\simeq (-\rmi)^2 \int_{q_0} S^f (q_L^+ ) \, \gamma_0 \, 
\left[\,  \int_{k_3} D_{n_\perp}^{ {\rm linear} } (q_3 - k_3)  \calM_{n_\perp, P_3}^{ff'} (k_3) \, \right] \gamma_{0}  S^{f'} (q_L^-) \,,
\end{equation}
where the one-dimensional potential is given by 
\begin{equation}
D_{n_\perp}^{ {\rm linear} } (q_3 )
 = \sigma \int_{|r_3| \gg |B|^{-1/2}} \rme^{\rmi q_3 r_3}  \sqrt{\, r_3^2 + \frac{\, 2n_\perp+1 \,}{\, |B| \,} \,} + O( \lqcd^4/|B|^2 ) 
\label{eq:1D_c}
\,.
\end{equation}
As in the harmonic oscillator model, 
the spectra are governed by the longitudinal dynamics until $n_\perp$ reaches $\sim |B|/\lqcd^2$, 
so that there are many low-energy states with the energies $\sim \lqcd$. 
While the square root induces the mixing between different harmonic oscillator bases, 
the overall tendencies are the same as in the result of the harmonic oscillator model. 
In fact, the above observation suggest that, as far as the above expansion is valid, 
the large degeneracy of the low energy bound states is 
a universal feature of long-range interaction models. 

\section{Meson resonance gas at finite temperature: percolation and chiral restoration}
\label{sec:HRG}

In the last sections 
we examined the properties of the low-energy bound states on the basis of the Bethe-Salpeter equations, 
and obtained the proper quantum numbers 
to specify their states. 
Especially, we discussed the difference between the neutral and charged mesons, 
and also the large degeneracy in the low energy domain. 
Without elaborating quantitative aspects of the meson structures, 
one can readily make a number of qualitative statements on the thermal partition function 
by using only the results obtained in the last sections. 

In this section we first examine the density of states of mesons, 
and then consider the thermodynamic partition function for the $q\bar{q}$ sector. 
We address how the percolation and chiral restoration proceed as temperature increases 
in the presence of strong magnetic fields.

\subsection{Density of states}

As a warming up, we begin with the review of the density of states for a quark, 
and then consider those for neutral and charged mesons. 
Below we concentrate on the density of states in the transverse directions 
as the continuous longitudinal spectrum is the same as in $B=0$ case. 
To count the number of a charged fermion state in magnetic fields, 
we will use a periodic lattice with the lattice spacing $a = 2\pi \Lambda_{ {\rm UV} }^{-1}$ 
and with the lengths $(L_1, L_2)$. These results provides 
the integral measures of the partition functions in the next subsection. 

\subsubsection{Density of states for single quarks}

Without magnetic fields, the total number of states for single fermions is 
as usual given by 
\begin{equation}
 \frac{\, L_1 L_2 \,}{a^2}  \sim \sum_{n_1}^{ L_1/a} ~\sum_{n_2}^{L_2/a} \sim 
\frac{\, L_1 L_2 \,}{ (2\pi)^2 } \int^{\luv} \! \rmd p_1 \int^{\luv} \! \rmd p_2  
\, ,
\end{equation}
%
This is nothing but the number of sites, and we converted it into the continuum expression. 
(We do not take into account non-integer part of $L/a$ and spins.) 
In the presence of magnetic fields, these states are divided into several LLs. 
In the Landau gauge, only momentum $p_2=2\pi n_2/L_2$ ($n_2$: integer) is conserved, 
so that a fermion state in the transverse dynamics is labeled by $p_2$ and 
the index of the Landau level, $n_L$. 
Also, since the location of the fermion is given by $R_1 = p_2/B$, 
this coordinate should be smaller than the length $L_1$, 
leading to the following condition on the maximum number of $n_2$: 
\begin{equation}
0 \le \frac{ 2\pi n_2}{|B| L_2} \le L_1 ~~ \rightarrow ~~ n_2^{ {\rm max} } \sim \frac{\, |B| L_1 L_2 \,}{ 2\pi } \,,
\end{equation}
which tells us the number of states for each LL. 
Then the sum of states can be written as 
\begin{equation}
 \frac{\, L_1 L_2 \,}{a^2}  ~\sim~ \sum_{n_L = 0}^{ 2\pi /|B| a^2 } \times 
\sum_{n_2= 0}^{ |B| L_1 L_2/ 2\pi }
~ \sim ~ \sum_{n_L = 0}^{ \Lambda_{ {\rm UV} }^2  /( 2\pi|B|) } ~\frac{L_2}{\, 2\pi \,} \int_0^{ |B| L_1} \! \rmd p_2 \,.
\end{equation}
%
To determine the maximum of $n_L$, we divided the total number of states $\sim L_1L_2/a^2$ by the number of states in each Landau level, $\sim |B| L_1L_2$. 
In most cases quantities, that we study, are independent of $p_2$, 
and we can carry out the integral over $p_2$. 
Dividing the number of states by the system volume $L_1 L_2$, 
we find the well-known degeneracy factor $|B|/2\pi$ for each LL. 
.

\subsubsection{Density of states for mesons}

Without magnetic fields, the total number of meson states are given by
\begin{equation}
 \left( \frac{\, L_1 L_2 \,}{a^2} \right)^2  
 ~ \sim ~ \sum_{n_1}^{ L_1/a} ~\sum_{n_2}^{L_2/a}  ~\sum_{n'_1}^{ L_1/a} ~\sum_{n'_2}^{L_2/a} 
 ~ \sim ~ \frac{\, L_1 L_2 \,}{ (2\pi)^2 } \int^{ \Lambda_{ {\rm UV} } }  \! \rmd \vP_\perp 
 \times \sum_{N_{b } =0 }^{L_1L_2/a^2} 
\, ,
\end{equation}
where we have reorganized the sum into the total and relative momenta, 
and $N_b$ labels bound states. 
Note that there is a maximum for the number of bound states, $N_b^{\rm max} = L_1 L_2/a^2$, reflecting 
the condition that the size of meson should be much smaller than the system size, 
or much larger than the lattice spacing. 

In the presence of magnetic fields, the sum is decomposed as
\begin{equation}
 \left( \frac{\, L_1 L_2 \,}{a^2} \right)^2  
 ~\sim~ \sum_{n_L = 0}^{ 2\pi /|B| a^2 } ~\sum_{n'_L = 0}^{ 2\pi /|B| a^2 } 
 \times \sum_{n_2= 0}^{ |B| L_1 L_2/ 2\pi } ~ \sum_{n'_2= 0}^{ |B| L_1 L_2/ 2\pi } \,.
\end{equation}
%
Now we focus on the meson states made only of LLLs. 
For such mesons with $n_L=n'_L=0$, the maximum number of states on the lattice is $\sim ( |B|L_1 L_2)^2$. 
As in the $B=0$ case, we can reorganize the summation of $n_2$ and $n_2'$ 
as the summation for conserved total momenta, $n^G_2 = n_2 + n'_2$, 
and relative momenta $\delta n_2 = n_2-n'_2$ as 
\begin{equation}
\left( \frac{ |B| L_1 L_2}{2\pi} \right)^2  ~\sim~ \sum_{n_2^G = 0}^{ |B| L_1 L_2/ 2\pi } \times \sum_{ \delta n_2 = 0}^{ |B| L_1 L_2/ 2\pi } 
~ \sim ~ \frac{L_2}{\, 2\pi \,} \int_0^{ |B| L_1} \! \rmd P_2 \times \sum_{ \delta n_2 = 0}^{ |B| L_1 L_2/ 2\pi } \,.
\end{equation}
Depending on whether meson states are neutral or charged, 
the summation over $\delta n_2$ is rearranged in two different ways below. 

For neutral mesons, we have seen that, in addition to $P_2$, 
one can define another conserved momentum that is called $P_1$. 
The momentum $P_1$ is also continuous and specifies the location of mesons in the $x_2$-direction 
as $P_2$ does in the $x_1$-direction. Therefore we have 
\begin{equation}
\left( \frac{\, |B| L_1 L_2 \,}{ 2\pi } \right)^2  
~ \sim ~ \frac{L_2}{\, 2\pi \,} \int_0^{ |B| L_1} \! \rmd P_2 \times  \frac{L_1}{\, 2\pi \,} \int_0^{ |B| L_2} \! \rmd P_1 
\, ~~~~({\rm neutral~mesons})
.
\end{equation}
On the other hand, the charged mesons have discrete spectra 
instead of the continuous $P_1$, so that the sum is expressed as 
\begin{equation}
\left( \frac{\, |B| L_1 L_2 \,}{ 2\pi } \right)^2  
~ \sim ~ \frac{L_2}{\, 2\pi \,} \int_0^{ |B| L_1} \! \rmd P_2 \times \sum_{ N_b = 0}^{ |B| L_1 L_2/ 2\pi } 
\, ~~~~({\rm charged~mesons})
,
\end{equation}
where $N_b$ labels the bound states as in the case without a magnetic field. 
These expressions will be used for the calculation of the partition function below.

\subsection{Mesonic partition function at finite temperature}

Finally we investigate the partition function at the temperature 
that is small enough to apply the dilute meson gas picture. 
At $B=0$, the non-interacting hadron resonance gas (HRG) picture reproduces 
the lattice data in a good accuracy until the temperature reaches the (pseudo)-critical temperature $T_c$. 
Around $T_c$, the thermally excited hadrons begin to overlap each other, 
and the interactions are no longer negligible; 
this is the regime where quarks and gluons start to emerge as the manifest 
microscopic degrees of freedom, 
leading to the quark-gluon plasma. We apply this established picture to 
the thermodynamics in the presence of large $B$.

%
%

Below we focus on the mesonic contribution to the pressure.
At the dilute regime such that the interactions are negligible, the partition function for mesons is written as
\begin{equation}
Z_M (T) \simeq \prod_n \prod_{P} \frac{1}{\, 1 - \rme^{ -E_n (P)/T } \,} \,,
\end{equation}
where $n$ and $P$ label a bound state and its total momentum, respectively. 
The corresponding pressure\footnote{In this paper by pressure we mean the negative of the free energy density. Precisely speaking,
there are two schemes to define the pressure depending on how one changes the volume of the system \cite{Bali:2014kia}. The first scheme fixes
the magnetic field density $B$ and the pressure is spatially isotropic, while the second scheme fixes the total number of magnetic flux leading to the spatial anisotropy in pressure \cite{Ferrer:2012wa}. In this paper we consider pressure in the first scheme.} is
\begin{equation}
\calP_M (T) \simeq  - T \sum_{n} \sum_P \ln \left(\, 1 - \rme^{-E_n(P)/T} \, \right) \,,
\end{equation}
where $E_n$ is the energy of a meson.

Now we use the specific forms of the density of states obtained in the previous subsection. 
Contributions from the neutral mesons made of the LLLs can be written as 
\begin{equation}
\calP^{\, {\rm neutral} }_M (T) \simeq  - T \sum_{f,f' } \delta_{ e_f, e_{f'} } \sum_{n_3} 
\int_{\vP_\perp} \int_{P_3} \ln \left(\, 1 - \rme^{-E_{n_3, \vP_\perp }^{ff'} (P_3)/T} \, \right) + \cdots\,,
\end{equation}
where the ellipses denote contributions from the interactions and the hLLs, 
and $\delta_{ e_f, e_{f'} }$ is unity for equal charges, otherwise zero. 
We should not take the spin sum because the flavor and the direction of spin is locked 
in the LLL. 
As discussed in the last subsection, the integrals with respect to the two transverse momenta 
have the symmetric form in case of the neutral mesons. 
On the other hand the charged mesons made of the LLLs provide the following contributions
\begin{equation}
\calP^{\, {\rm charged} }_M (T) \simeq  - T \sum_{f,f' } \left( 1 - \delta_{ e_f, e_{f'} } \right) 
\sum_{n_3} \sum_{ n_\perp } c_{ff'} \frac{\,  |B| \,}{\, 2\pi \,}\int_{P_3}  \ln \left(\, 1 - \rme^{-E_{ n_3, n_\perp }^{ff'} (P_3)/T} \, \right) + \cdots \,,
\end{equation}
where $n_\perp$ labels the bound state in the transverse dynamics. 
Note that we could integrate over $P_2$ and get the factor $c_{ff'} |B|/2\pi$ 
since the spectrum is independent of $P_2$: 
$c_{ff'}$ is some number associated with the electric charges of the flavors $f$ and $f'$.

Following from the analyses of the Schwinger-Dyson and Bethe-Salpeter equations, 
one can deduce approximate forms of the low energy meson spectra 
for long-range interaction models. 
We define the mass of ground state meson at each quantum number as
\begin{equation}
M^{ {\rm neutral} }_{n_3} \equiv E_{n_3, \vP_\perp = \vec{0}  }^{ff'} (P_3=0) \,,
~~~~~~
M^{ {\rm charged} }_{n_3} \equiv E_{n_3, n_\perp =0  }^{ff'} (P_3=0) \,, 
\end{equation}
where $P_3=0$ and the quantum numbers for the transverse direction take the minimal number, e.g., $P_2=0$. 
While we have not shown explicit numerical solutions of the energy levels on the right-hand sides, 
we showed that with these quantum numbers 
the dynamics is governed by the ($1+1$)-dimensional dynamics in the $x_L$-direction, 
and the meson spectra are independent of the magnitude of $B$. 
Now we increase $P_3$. Since the Poincare invariance should be maintained in the $x_L$-direction,
the spectra are 
\begin{equation}
E_{n_3, \vP_\perp = \vec{0}  }^{ff'} (P_3) = \sqrt{ \left(M^{ {\rm neutral} }_{n_3} \right)^2 + P_3^2 \, } \,,
~~~~~~
E_{n_3, n_\perp =0}^{ff'} (P_3)
\simeq 
\sqrt{ \left(M^{ {\rm charged} }_{n_3} \right)^2 + P_3^2 \, } \,.
\end{equation}
Next we further increase the quantum numbers associated with the transverse dynamics.
There is no kinetic term in the transverse directions, so that the contributions to the spectra 
arise only from the potential energy which we have examined approximately 
in Eqs.~(\ref{eq:1D_n}) and (\ref{eq:1D_c}). 
Including them as corrections to the above spectra, we arrive at
\begin{equation}
E_{n_3, \vP_\perp }^{ff'} (P_3) 
\simeq 
\sqrt{ \left(M^{ {\rm neutral} }_{n_3} \right)^2 + P_3^2 \, }
+ c_1 \lqcd^3 \, \frac{\, P_\perp^2 \,}{\, |B|^2 \,} + \cdots \,,~~~~~~~~~ ({\rm for~neutral~mesons})
\end{equation}
and 
\begin{equation}
E_{n_3, n_\perp }^{ff'} (P_3)
\simeq 
\sqrt{ \left(M^{ {\rm charged} }_{n_3} \right)^2 + P_3^2 \, }
+ c_2 \lqcd^3 \, \frac{\, n_\perp \,}{\, |B| \,} + \cdots \,, ~~~~~~~~~ ({\rm for~charged~mesons})
\end{equation}
where $c_1$ and $c_2$ are dimensionless 
constant\footnote{Strictly speaking, 
$c_1$ and $c_2$ can be functions of $P_3$
since our evaluation is based on the expansion of $\sim 1/\la r_3 \ra$ and $\la r_3\ra$ can depend on 
the Lorentz boost. But as far as $P_3$ is small we can treat $c_1$ and $c_2$ as constants 
and this is the situation we will discuss in the following.}. 

\begin{figure}[!tb]
\begin{center}
\includegraphics[width = 0.9\textwidth]{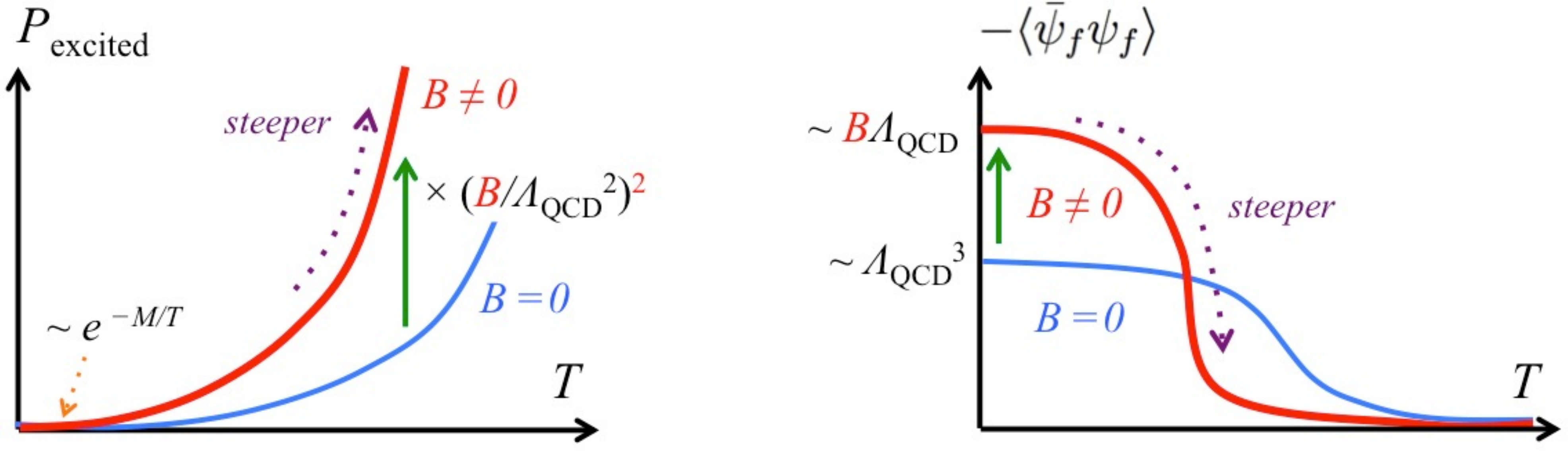}
\end{center}
\vspace{0.cm}
\caption{
\footnotesize{(Left) Contributions of the meson excitations to the finite $T$ pressure at $B=0$ and $B\neq 0$. 
$P_{ {\rm excited} }$ at $|eB| \gg \lqcd^2$ is bigger than the pressure at $B=0$ 
roughly by a factor of $(|eB|/\lqcd^2)^2$. 
The pressure starts to contribute significantly when $T$ reaches $\sim M \sim \lqcd $. (Right) The finite $T$ chiral condensate, $-\la \bar{\psi}_f \psi_f \ra_{T}$. At $T=0$, the (absolute value of) condensate grows linearly as $B$ increases. On the other hand, as $T$ increases, 
the condensate at $B\neq 0$ decreases more rapidly than the $B=0$ condensate 
because of the hadron resonances which emerge more radically at finite $B$, giving the chiral density of $\sim B^2$ with the opposite sign to the vacuum chiral condensate.
}
\vspace{0.3cm}
}\label{fig:HRG}
\end{figure}

At temperature lower than the meson masses such that $T\ll M_{n_3}$, 
one can use the expansion $\ln (1-\rme^{-x} ) \simeq -\rme^{-x}$ to obtain an analytic form of the pressure, 
and then finds that both of the contributions from neutral and charged mesons 
behave as
\begin{equation}
\calP_n (T)  \sim  T^2  \frac{\, |B|^2 \,}{\, \lqcd^3 \,} \int_{P_3} \rme^{-\sqrt{M_{n_{3}}^2 + P_3^2 \,} /T } 
 \sim 
T^2  \frac{\, |B|^2 \,}{\, \lqcd^3 \,} 
\times (M_{n_{3}} T)^{1/2} \, \rme^{-M_{n_{3}}/T} + \cdots  \,,
\end{equation}
where we used the non-relativistic approximation to get the final expression, 
and a factor of $|B|^2$ is originated from the transverse phase space volume, 
namely either the integration of $P_2$ or summation of $n_2$. 
Since a lot of low energy states is involved in the transverse dynamics, 
we get a big phase factor $\sim |B|^2/\lqcd^2$. 
The above should be compared with the pressure at $B=0$ generated by mesons with mass $M_n'$,
\begin{equation}
\calP^{B=0}_n (T)  \sim  T \int_{\vP} \rme^{-\sqrt{M_n^{'2} + \vP^2 \,} /T } 
 \sim 
T  \times (M'_n T)^{3/2} \, \rme^{-M'_n/T} + \cdots  \,.
\end{equation}
In both $B=0$ and large $B$ cases, the pressures have the same temperature dependences. 
However, clearly, 
the magnetic-field dependence of the rest mass $M_n$ plays a crucial role to determine 
whether the effects of strong magnetic fields act to enhance or suppress the pressure $\calP$. 
Recall that, for long-range interactions, the magnetic-field dependence decouples 
from $M_n$ at very large $|B|$, and $M_n$ stays around $\sim \lqcd$. 
In this regime, the mesonic pressure keeps growing as $|B|$ increases. 
Then the system reaches the percolation region at smaller $T$ than in the $B=0$ case; 
namely, thermally excited hadrons overlap each other, 
so that gluons propagate from one to another hadron as if they are liberated from hadrons.
The schematic picture is shown in Fig.~\ref{fig:HRG} (left).

This percolation picture of meson gas can be also combined with the chiral restoration at finite $T$. 
The chiral condensate with the flavor $f$ can be calculated by taking a derivative of $\calP$ 
with respect to the current quark mass as 
\begin{equation}
- \la \bar{\psi}_f \psi_f \ra_T 
= \frac{\, \partial \calP_{ {\rm vac} } \,}{\, \partial m_f \,} + \frac{\, \partial \calP_{ {\rm excited} } \,}{\, \partial m_f \,}
\simeq  - \la \bar{\psi}_f \psi_f \ra_{T=0} - \sum_n \sum_P \frac{\, \partial E_n(P) \,} { \partial m_f} \frac{1}{\, \rme^{E_n(P)/T } - 1 \,}  \,,
\end{equation}
where $\calP_{ {\rm vac} }$ and $\calP_{ {\rm excited} }$ are the contributions to pressure from the vacuum 
and thermally excited states, respectively, and $\partial E_n/\partial m_f$ is the sigma term of a meson. 
Note that the vacuum value of the condensate is negative $\la \bar{\psi} \psi \ra_{T=0} <0$, 
and also that the sigma term should be positive, 
giving a competing contribution 
to the vacuum contribution. 
Therefore, with more and more meson excitations, 
the chiral condensate is strongly diminished by the meson gas contributions. 

We have already seen that the meson spectra at large $B$ stay as small as those in $B=0$, 
so that they are easily excited at finite temperature. 
While at very low temperature the vacuum contribution grows by the magnetic catalysis, 
the mesonic contribution at large $B$ grows faster than the vacuum contribution 
once they are activated in finite $T$, thanks to the large phase space volume in the transverse dynamics. 
Therefore, the relative magnitude of the growth rates is {\it inversed}, 
leading to the dissociation of the chiral condensate at lower $T$ 
compared to the $B=0$ case [see Fig.\ref{fig:HRG} (right)]. 
The percolation and chiral restoration are intimately connected; 
the enhanced number of hadronic excitations provides 
a simple explanation of the inverse magnetic catalysis phenomenon. 

\section{Summary}
\label{sec:summary}

In this first part of a series of papers, we have provided heuristic arguments on the meson structures and spectra. Our analyses were specialized to models of long-range interactions such as the confining linear rising potential or harmonic oscillator. The key assumption was that these long-range contributions dominate over the short-range ones, although presumably this picture might oversimplify the reality. Nevertheless, such studies can cover 
aspects of the interplay between non-perturbative QCD interactions and strong magnetic fields 
which have not been addressed in the previous studies 
since they primarily rely on effective models with short-range interactions.

On the basis of our assumption of the IR-dominant interaction, 
we systematically studied the parametric behaviors of the various quantities. 
These systematic results, covering from the constituent quark mass 
to the meson spectra and their thermodynamics, 
were missing in the conventional studies, and provides a consistent guideline 
to understand the lattice results. Especially, we argued an importance of 
the long-range interaction for understanding the lattice results 
and the reason why the conventional effective models of short-range interactions 
have not been able to explain them. 

We note that, while there were several studies treating either the Schwinger-Dyson or Bethe-Salpeter equations at finite $B$ separately, our work is the first attempt to treat the quark self-energies and meson spectra in a consistent way. The consistent treatments are essential when we consider quantities related to the spontaneous symmetry breaking, such as spectra of the Nambu-Goldstone bosons.

As found in our previous studies, the quark self-energies and mass gap are $B$-independent for interactions with the IR dominance. This behavior is in favor of explaining the linear $B$-dependence of the chiral condensate, which has been observed in the lattice simulations. With the same mechanism, it follows that the meson states at low energy are also nearly $B$-independent. Moreover, we found interesting $B$-dependence of the transverse momenta, which is essential to count the density of states of mesons at low energy. With this $B$-dependence of the meson spectra, 
it was shown that at low temperature, the pressure of meson gas parametrically grows as $\sim B^2$ as $|B|$ increases. 
This behavior was commonly found in contributions of both neutral and charged mesons. 
In particular, contributions from neutral mesons are drastically different from those at small $|B|$; 
strong magnetic fields can intrude a neutral composite meson, 
causing a strong structure change and modification of the spectrum. 

The percolation and chiral restoration were discussed within the hadron resonance gas picture. Two phenomena are intimately connected by the overlap of thermally excited hadrons; at larger $B$, such overlap occurs at lower $T$, because of the enhanced phase space due to the magnetic fields.

Finally we close our discussions by repeating the importance of the IR gluons, since their impacts on the QCD phenomenology are so distinct from those based on perturbative gluons or short-range interactions.
The examples can be found for QCD at finite density such as the quarkyonic chiral spirals~\cite{chiralspirals,Ferrer:2012zq}, at finite temperature such as the improved equation of state~\cite{EoS} and the massless mode which induces positivity violation~\cite{Su:2014rma}, 
and also in external magnetic fields such as the quark mass gap which resolves the puzzle posed by the lattice results~\cite{Kojo:2012js,Watson:2013ghq}. Furthermore, the application of such an approach to heavy-ion phenomenology has shown signatures of a strongly coupled quark-gluon plasma~\cite{pheno}. All these studies so far have shown significant importance of the role of long-range interactions/confinement effects in various aspects of QCD under extreme conditions, ranging from static to dynamic properties. It is thus very encouraging to keep exploring along this direction, which will certainly improve our understanding of the QCD phase transition.

In the second part, we will address more quantitative aspects, using models of various types of interactions.

\section*{Acknowledgments}
This work was supported by NSF Grants PHY09-69790 and PHY13-05891 (T.K.), 
JSPS Grants-in-Aid No.~25287066 (K.H.), and 
the Helmholtz International Center for FAIR within the framework of 
the LOEWE program launched by the State of Hesse (N.S.).

\appendix
\section{Dimensionally reduced operators}
\label{sec:app}

\label{sec:locking}

In this paper we consider only the strong field regime 
where the LLL approximation is the proper starting point. 
In such a regime, usual classifications of states, in terms of the angular momentum or isospin, 
are not as useful as in the $B=0$ case. 
On the other hand, we are also interested in how meson states at $B=0$ 
are connected to those at large $B$. To observe the relation between these two limiting cases, 
we investigate how the meson interpolating fields at $B=0$ reduce to 
the (1+1) dimensional fields in the LLLs.
The good summary of the algebra can be also found in Ref.\cite{Basar:2010zd}. 

\subsection{Quantum numbers of the LLL fields}

Let us begin with the $\gamma_5$-operator, 
\begin{equation}
\gamma_5 = \gamma_0 \gamma_1 \gamma_2 \gamma_3 = (- \rmi \gamma_0 \gamma_3) (\rmi \gamma_1 \gamma_2) \equiv \Gamma_5 \sigma_3 \,, 
\end{equation}
where $\sigma_3$ is the third component of the spin operator, 
and $\Gamma_5 \equiv -\rmi \gamma_0 \gamma_3$ can be regarded 
as the two-dimensional $\gamma_5^{ {\rm 2D} }$ which characterizes 
the moving directions in the $x_3$-direction. Using $\Gamma_5$, 
we can define the right- and left-movers whose momenta are oriented in  
the $+x_3$ and $-x_3$ directions, respectively: 
\begin{equation}
\frac{\, 1\pm \Gamma_5 \,}{2}\, \psi \equiv \psi_{r,\,l} \,.
\end{equation}
We use the indices ($r,l$) to label the moving directions (or two-dimensional chirality), 
and distinguish them from the four-dimensional chirality ($R,L$). 
The relation between $(R,L)$ and $(r,l)$ depends on 
the spin directions of fields. We find the relations for spin up and down states, respectively, as 
\begin{equation}
( \psi_\up )_{R, L}  = \frac{1 \pm \gamma_5}{2}\,  \psi_\up  =  \frac{1 \pm \Gamma_5}{2} \,  \psi_\up =  ( \psi_\up )_{r,\, l} \,,
~~~~
( \psi_\down )_{R, L}  = \frac{1 \pm \gamma_5}{2}\,  \psi_\down  =  \frac{1 \mp \Gamma_5}{2} \,  \psi_\down =  ( \psi_\down )_{l,\, r} \,.
\end{equation}
For the spin down states, the chiralities in four- and two-dimension appears to be opposite.

Now let $B>0$ and consider the LLL fields. The spin-flavor contents are $\chi^f = (u_\up, d_\down, \cdots)$. 
As in the above, we find\footnote{These expressions are useful to intuitively understand the charge separation effect in the presence of chiral imbalance. With an excess of the right-handed chirality, we have more $u_\up$ quark moving in the $x_3$-direction, and more $d_\down$ quark moving in the $-x_3$-direction.
}
\begin{equation}
\chi^f_R = (  u^r_{\up} \,, d^l_\down \,, \cdots) \,,~~~~~~ \chi^f_L = (  u^l_\up \,, d^r_\down \,, \cdots) \,.
\end{equation}
When considering $B<0$, one should flip the direction of spin.

\subsection{Quark bilinear operators}

Next we map the four-dimensional quark bilinear operators from four to two space-time dimensions. 
In doing this we will keep only the LLL fields. 
The scalar and pseudo-scalar operators which can be made of the LLL fields are listed as 
\begin{equation}
\left(\, \bar{u} u\,,  \bar{d} d \, \right)_{ {\rm LLL} }  
	\, \sim \, \left(\, \bar{\chi}_u \chi_u \,,\, \bar{\chi}_d \chi_d \, \right) \,,~~~~
\left( \, \bar{u} \rmi \gamma_5 u \,, \, \bar{d} \rmi \gamma_5 d \, \right)_{ {\rm LLL} } 
	\, \sim \, 
	\left(\, \bar{\chi}_u \rmi \Gamma_5 \chi_u \,, \,  - \bar{\chi}_d \rmi \Gamma_5 \chi_d \, \right)
\, .
\end{equation}
Operators such as a $\bar{u}d$ necessarily contain the hLL fields, 
so that such combinations were omitted here. 
These expressions imply that the scalar and pseudo-scalar mesons are light when they are neutral, e.g., $\pi^0$, 
and are as heavy as $\sqrt{|eB|}$ when they are charged, e.g., $\pi^\pm $. 
Note that in the pseudo-scalar channel the sign of the operator flips for $d$-quarks. 
Due to this sign flip, the isosinglet (isovector) operators made of the LLLs in 4D 
behave as isovector (isosinglet) in 2D. This observation becomes important 
when one considers the annihilation of a quark and anti-quark in the LLLs 
into purely gluonic configurations, such as instantons.

For vector currents, there are two types of operators that couple to low energy states. 
The first type is composed of like-charge fields 
\begin{equation}
\left(\, \bar{u} \gamma_L u\,,  \bar{d} \gamma_L d \, \right)_{ {\rm LLL} }  
	\, \sim \, \left(\, \bar{\chi}_u \gamma_L \chi_u \,,\, \bar{\chi}_d \gamma_L \chi_d \, \right) \,,~~~~
\end{equation}
where $\gamma_L$ preserves the spin of fields. 
For instance, it couples $\bar{u}_\up$ to $u_\up$, or $\bar{d}_\down$ to $s_\down$, etc. 
The expression implies that vector mesons with $s_z=0$ become light 
when they are neutral, e.g., $\rho^0_{s_z=0}$ and $\omega_{s_z=0}$, 
but becomes heavy when they are charged, e.g., $\rho^\pm_{s_z=0}$. 
The second type is composed of unlike-charge fields 
\begin{equation}
\left(\, \bar{u} \gamma_\perp d\,,  \bar{d} \gamma_\perp u \, \right)_{ {\rm LLL} }  
	\, \sim \, \left(\, \bar{\chi}_u \gamma_\perp \chi_d \,,\, \bar{\chi}_d \gamma_\perp \chi_u \, \right) \,,~~~~
\end{equation}
where $\gamma_\perp$ flips the spin of fields and thereby can couple the LLL fields with different flavors. 
In this respect, $\gamma_\perp$-matrices may be regarded as "flavor"-matrices for the LLL dynamics. 
Note also that $u$ and $d$ in the LLL have the opposite two-dimensional chirality, 
so that $\bar{\chi}_{u} \gamma_\perp \chi_{d}$ and $\bar{\chi}_{d} \gamma_\perp \chi_{u}$ 
behave as 2D scalar operators, rather than vector ones. 
Therefore 4D operators producing mesons with $s_z = \pm 1$ states 
end up with scalar mesons in 2D at large $B$. 
The expression implies the spin-charge locked states which are realized in such a way 
that $\rho^\pm_{s_z=\pm1}$ are light, while $\rho^\pm_{s_z= \mp 1} $, $\rho^0_{s_z=\pm1} $, etc. are heavy. 

Next we consider axial-vector currents. The longitudinal components can be related to the vector currents as
\begin{equation}
\left(\, \bar{u} \gamma_L \gamma_5 u\,,  \bar{d} \gamma_L \gamma_5 d \, \right)_{ {\rm LLL} }  
 	\, \sim \, \left(\, \bar{\chi}_u \gamma_L \Gamma_5 \chi_u \,,\, - \bar{\chi}_d \gamma_L \Gamma_5 \chi_d \, \right) 
 	\, \sim \, (-\rmi) \, \epsilon_{L L'} \left(\, \bar{\chi}_u \gamma_{L'} \chi_u \,,\, - \bar{\chi}_d \gamma_{L'}  \chi_d \, \right) 	
	\,.~~~~
\end{equation}
Therefore the 4D axial-vector and vector currents couple to the same 2D states. Since the 4D axial-vector currents also couple to the Nambu-Goldstone bosons, at large $|B|$ the correlators of axial-vector, vector, and pseudo-scalar show the same low energy behaviors governed by the LLLs.
The transverse components are
\begin{equation}
\left(\, \bar{u} \gamma_\perp \gamma_5 d\,,  \bar{d} \gamma_\perp \gamma_5 u \, \right)_{ {\rm LLL} }  
 	\, \sim \, \left(\, \bar{\chi}_u \gamma_\perp \Gamma_5 \chi_d \,,\, - \bar{\chi}_d \gamma_\perp \Gamma_5 \chi_u \, \right) 
	\,,~~~~
\end{equation}
which can be regarded as flavor off-diagonal 2D pseudo-scalar operators. 

For tensor operators, there are three kinds of operators. 
The first two operators do not mix up different spin states 
and can be reduced to ($\sigma_{\mu \nu}= \rmi [\gamma_\mu,\gamma_\nu]/2$)
\begin{equation}
\left(\, \bar{u} \sigma_{LL'} u\,,  \bar{d} \sigma_{LL'} d \, \right)_{ {\rm LLL} }  
 	\, \sim \,  - \epsilon_{LL'} \left(\, \bar{\chi}_u \Gamma_5 \chi_u \,,\,  \bar{\chi}_d \Gamma_5 \chi_d \, \right) \,,
\end{equation}
and 
\begin{equation}
\left(\, \bar{u} \sigma_{\perp \perp'} u\,,  \bar{d} \sigma_{\perp \perp'} d \, \right)_{ {\rm LLL} }    
 	\, \sim \,  \epsilon_{\perp \perp'} \left(\, \bar{\chi}_u  \chi_u \,,\,  -\bar{\chi}_d \chi_d \, \right) \,.
\end{equation}
The third one mixes spins (or flavors), 
\begin{equation}
\left(\, \bar{u} \sigma_{L \perp} d\,,  \bar{d} \sigma_{L\perp} u \, \right)_{ {\rm LLL} }    
 	\, \sim \,   \rmi  \left(\, \bar{\chi}_u \gamma_L \gamma_\perp \chi_d \,,\,  \bar{\chi}_d \gamma_L \gamma_\perp \chi_u \, \right) \,,
\end{equation}
which can be regarded as flavor off-diagonal 2D vector operators. 

In summary, many 4D operators, that couple to different meson states, 
show the same asymptotic behaviors, and can be reduced to 2D operators. 
In particular, operators having very different structures in four dimensions 
can look the same in the language of 2D operators made of the LLLs. 
For instance, we have seen that the axial-vector, vector, and pseudo-scalar operators in 4D 
couple to the same pseudo-scalar mesons in 2D. 
This implies that in strong magnetic fields these correlators at a long distance scale 
are saturated by the same asymptotic states.



\begin{thebibliography}{00}
%
%
\bibitem{Miransky:2015ava}
  V.~A.~Miransky and I.~A.~Shovkovy,
  Phys.\ Rept.\  {\bf 576} (2015) 1
  [arXiv:1503.00732 [hep-ph]];
J.~O.~Andersen, W.~R.~Naylor and A.~Tranberg,
  arXiv:1411.7176 [hep-ph];
    D.~Kharzeev, K.~Landsteiner, A.~Schmitt and H.~U.~Yee,
  Lect.\ Notes Phys.\  {\bf 871} (2013) pp.1.


\bibitem{Kharzeev:2007jp}
  D.~E.~Kharzeev, L.~D.~McLerran and H.~J.~Warringa,
  Nucl.\ Phys.\ A {\bf 803} (2008) 227
  [arXiv:0711.0950 [hep-ph]];
  K.~Fukushima, D.~E.~Kharzeev and H.~J.~Warringa,
  Phys.\ Rev.\ D {\bf 78} (2008) 074033
  [arXiv:0808.3382 [hep-ph]];
  K.~Tuchin,
  Adv.\ High Energy Phys.\  {\bf 2013} (2013) 490495
  [arXiv:1301.0099];
  L.~McLerran and V.~Skokov,
  Nucl.\ Phys.\ A {\bf 929} (2014) 184
  [arXiv:1305.0774 [hep-ph]].
  
\bibitem{Skokov:2009qp}
  V.~Skokov, A.~Y.~Illarionov and V.~Toneev,
  Int.\ J.\ Mod.\ Phys.\ A {\bf 24} (2009) 5925
  [arXiv:0907.1396 [nucl-th]];
  A.~Bzdak and V.~Skokov,
  Phys.\ Lett.\ B {\bf 710} (2012) 171
  [arXiv:1111.1949 [hep-ph]];
  V.~Voronyuk, V.~D.~Toneev, W.~Cassing, E.~L.~Bratkovskaya, V.~P.~Konchakovski and S.~A.~Voloshin,
  Phys.\ Rev.\ C {\bf 83} (2011) 054911
  [arXiv:1103.4239 [nucl-th]];
  W.~T.~Deng and X.~G.~Huang,
  Phys.\ Rev.\ C {\bf 85} (2012) 044907
  [arXiv:1201.5108 [nucl-th]].
  
\bibitem{Vachaspati:1991nm}
  T.~Vachaspati,
  Phys.\ Lett.\ B {\bf 265} (1991) 258;
  K.~Enqvist and P.~Olesen,
  Phys.\ Lett.\ B {\bf 319} (1993) 178
  [hep-ph/9308270].

%
%
\bibitem{Buividovich:2008wf}
  P.~V.~Buividovich, M.~N.~Chernodub, E.~V.~Luschevskaya and M.~I.~Polikarpov,
  Phys.\ Lett.\ B {\bf 682} (2010) 484
  [arXiv:0812.1740 [hep-lat]];
  M.~D'Elia and F.~Negro,
  Phys.\ Rev.\ D {\bf 83} (2011) 114028
  [arXiv:1103.2080 [hep-lat]];
  G.~S.~Bali, F.~Bruckmann, G.~Endr${\rm \ddot{o} }$di, Z.~Fodor, S.~D.~Katz and A.~Sch${\rm \ddot{a} }$fer,
  Phys.\ Rev.\ D {\bf 86} (2012) 071502
  [arXiv:1206.4205 [hep-lat]].
  %

\bibitem{Bali:2011qj}
  G.~S.~Bali, F.~Bruckmann, G.~Endr${\rm \ddot{o} }$di, Z.~Fodor, S.~D.~Katz, S.~Krieg, A.~Sch${\rm \ddot{a}}$fer and K.~K.~Szabo,
  JHEP {\bf 1202} (2012) 044
  [arXiv:1111.4956 [hep-lat]].
  
\bibitem{Bornyakov:2013eya}
  V.~G.~Bornyakov, P.~V.~Buividovich, N.~Cundy, O.~A.~Kochetkov and A.~Sch${\rm \ddot{a}}$fer,
  Phys.\ Rev.\ D {\bf 90} (2014) 3,  034501
  [arXiv:1312.5628 [hep-lat]].

\bibitem{Bali:2013esa}
  G.~S.~Bali, F.~Bruckmann, G.~Endr${\rm \ddot{o} }$di, F.~Gruber and A.~Sch${\rm \ddot{a}}$fer,
  JHEP {\bf 1304} (2013) 130
  [arXiv:1303.1328 [hep-lat]].
\bibitem{Bonati:2014ksa}
  C.~Bonati, M.~D'Elia, M.~Mariti, M.~Mesiti, F.~Negro and F.~Sanfilippo,
  Phys.\ Rev.\ D {\bf 89} (2014) 11,  114502
  [arXiv:1403.6094 [hep-lat]].
  
  
  \bibitem{Gusynin:1994xp} 
  V.~P.~Gusynin, V.~A.~Miransky and I.~A.~Shovkovy,
  Phys.\ Lett.\ B {\bf 349}, 477 (1995)
  [hep-ph/9412257].

\bibitem{Fukushima:2011jc}
  K.~Fukushima,
  J.\ Phys.\ G {\bf 39} (2012) 013101
  [arXiv:1108.2939 [hep-ph]].
  
\bibitem{Kojo:2012js}
  T.~Kojo and N.~Su,
  Phys.\ Lett.\ B {\bf 720} (2013) 192
  [arXiv:1211.7318 [hep-ph]];
  {\it ibid.}
  Nucl.\ Phys.\ A {\bf 931} (2014) 763
  [arXiv:1407.7925 [hep-ph]].
  
 
 \bibitem{Hattori:2015hka} 
  K.~Hattori, K.~Itakura, S.~Ozaki and S.~Yasui,
  Phys.\ Rev.\ D {\bf 92}, no. 6, 065003 (2015)
  [arXiv:1504.07619 [hep-ph]].

\bibitem{Ozaki:2015sya} 
  S.~Ozaki, K.~Itakura and Y.~Kuramoto,
  arXiv:1509.06966 [hep-ph].
  

  

\bibitem{Shovkovy:2012zn}
  I.~A.~Shovkovy,
  Lect.\ Notes Phys.\  {\bf 871} (2013) 13
  [arXiv:1207.5081 [hep-ph]];
  H.~Suganuma and T.~Tatsumi,
  Annals Phys.\  {\bf 208} (1991) 470;
T.~D.~Cohen, D.~A.~McGady and E.~S.~Werbos,
  Phys.\ Rev.\ C {\bf 76} (2007) 055201
  [arXiv:0706.3208 [hep-ph]];
    R.~Gatto and M.~Ruggieri,
  Phys.\ Rev.\ D {\bf 83} (2011) 034016
  [arXiv:1012.1291 [hep-ph]];
  J.~O.~Andersen,
  Phys.\ Rev.\ D {\bf 86} (2012) 025020
  [arXiv:1202.2051 [hep-ph]];
   J.~O.~Andersen,
  JHEP {\bf 1210} (2012) 005
  [arXiv:1205.6978 [hep-ph]].

\bibitem{Endrodi:2015oba}
  G.~Endrodi,
  JHEP {\bf 1507} (2015) 173
  [arXiv:1504.08280 [hep-lat]].
  

\bibitem{Gatto:2010pt}
  R.~Gatto and M.~Ruggieri,
  Phys.\ Rev.\ D {\bf 83} (2011) 034016
  [arXiv:1012.1291 [hep-ph]];
  A.~J.~Mizher, M.~N.~Chernodub and E.~S.~Fraga,
  Phys.\ Rev.\ D {\bf 82} (2010) 105016
  [arXiv:1004.2712 [hep-ph]];
  K.~Kashiwa,
  Phys.\ Rev.\ D {\bf 83} (2011) 117901
  [arXiv:1104.5167 [hep-ph]].



\bibitem{Fukushima:2012kc}
  K.~Fukushima and Y.~Hidaka,
  Phys.\ Rev.\ Lett.\  {\bf 110} (2013) 3,  031601
  [arXiv:1209.1319 [hep-ph]].

\bibitem{Chao:2013qpa}
  J.~Chao, P.~Chu and M.~Huang,
  Phys.\ Rev.\ D {\bf 88} (2013) 054009
  [arXiv:1305.1100 [hep-ph]].
  L.~Yu, H.~Liu and M.~Huang,
  Phys.\ Rev.\ D {\bf 90} (2014) 7,  074009
  [arXiv:1404.6969 [hep-ph]];
  L.~Yu, J.~Van Doorsselaere and M.~Huang,
  Phys.\ Rev.\ D {\bf 91} (2015) 7,  074011
  [arXiv:1411.7552 [hep-ph]].
\bibitem{Feng:2014bpa}
  B.~Feng, D.~f.~Hou and H.~c.~Ren,
  Phys.\ Rev.\ D {\bf 92} (2015) 6,  065011
  [arXiv:1412.1647 [cond-mat.quant-gas]].

\bibitem{Bruckmann:2013oba}
  F.~Bruckmann, G.~Endr${\rm \ddot{o} }$di and T.~G.~Kovacs,
  JHEP {\bf 1304} (2013) 112
  [arXiv:1303.3972 [hep-lat]].

\bibitem{Ferrer:2014qka}
  E.~J.~Ferrer, V.~de la Incera and X.~J.~Wen,
  Phys.\ Rev.\ D {\bf 91} (2015) 5,  054006
  [arXiv:1407.3503 [nucl-th]].

\bibitem{Farias:2014eca}
  R.~L.~S.~Farias, K.~P.~Gomes, G.~I.~Krein and M.~B.~Pinto,
  Phys.\ Rev.\ C {\bf 90} (2014) 2,  025203
  [arXiv:1404.3931 [hep-ph]].
\bibitem{Ferreira:2014kpa}
  M.~Ferreira, P.~Costa, O.~Lourenço, T.~Frederico and C.~Providência,
  Phys.\ Rev.\ D {\bf 89} (2014) 11,  116011
  [arXiv:1404.5577 [hep-ph]].

\bibitem{Mueller:2015fka}
  N.~Mueller and J.~M.~Pawlowski,
  Phys.\ Rev.\ D {\bf 91} (2015) 11,  116010
  [arXiv:1502.08011 [hep-ph]].


\bibitem{Skokov:2011ib}
  V.~Skokov,
  Phys.\ Rev.\ D {\bf 85} (2012) 034026
  [arXiv:1112.5137 [hep-ph]].
\bibitem{Fukushima:2012xw}
  K.~Fukushima and J.~M.~Pawlowski,
  Phys.\ Rev.\ D {\bf 86} (2012) 076013
  [arXiv:1203.4330 [hep-ph]].
\bibitem{Kamikado:2013pya}
  K.~Kamikado and T.~Kanazawa,
  JHEP {\bf 1403} (2014) 009
  [arXiv:1312.3124 [hep-ph]];
{\it ibid.}
  JHEP {\bf 1501} (2015) 129
  [arXiv:1410.6253 [hep-ph]].

\bibitem{Ozaki:2013sfa}
  S.~Ozaki,
  Phys.\ Rev.\ D {\bf 89} (2014) 5,  054022
  [arXiv:1311.3137 [hep-ph]];
  S.~Ozaki, T.~Arai, K.~Hattori and K.~Itakura,
  Phys.\ Rev.\ D {\bf 92} (2015) 1,  016002
  [arXiv:1504.07532 [hep-ph]].

\bibitem{Watson:2013ghq}
  P.~Watson and H.~Reinhardt,
  Phys.\ Rev.\ D {\bf 89} (2014) 4,  045008
  [arXiv:1310.6050 [hep-ph]];
  N.~Mueller, J.~A.~Bonnet and C.~S.~Fischer,
  Phys.\ Rev.\ D {\bf 89} (2014) 9,  094023
  [arXiv:1401.1647 [hep-ph]].
  

\bibitem{'tHooft:1974hx}
  G.~'t Hooft,
  Nucl.\ Phys.\ B {\bf 75} (1974) 461;
  C.~G.~Callan, Jr., N.~Coote and D.~J.~Gross,
  Phys.\ Rev.\ D {\bf 13} (1976) 1649;
  I.~Bars and M.~B.~Green,
  Phys.\ Rev.\ D {\bf 17} (1978) 537.


\bibitem{Chernodub:2011gs}
  M.~N.~Chernodub, J.~Van Doorsselaere and H.~Verschelde,
  Phys.\ Rev.\ D {\bf 85} (2012) 045002
  [arXiv:1111.4401 [hep-ph]].
  


\bibitem{Simonov:2012if}
  Y.~A.~Simonov, B.~O.~Kerbikov and M.~A.~Andreichikov,
  arXiv:1210.0227 [hep-ph].
  M.~A.~Andreichikov, B.~O.~Kerbikov, V.~D.~Orlovsky and Y.~A.~Simonov,
  Phys.\ Rev.\ D {\bf 87} (2013) 9,  094029
  [arXiv:1304.2533 [hep-ph]].

\bibitem{Alford:2013jva}
  J.~Alford and M.~Strickland,
  Phys.\ Rev.\ D {\bf 88} (2013) 105017
  [arXiv:1309.3003 [hep-ph]].


\bibitem{Taya:2014nha}
  H.~Taya,
  Phys.\ Rev.\ D {\bf 92} (2015) 1,  014038
  [arXiv:1412.6877 [hep-ph]].
  
    \bibitem{Gub} S. Cho, K. Hattori, S. H. Lee, K. Morita, and S. Ozaki, 
Phys. Rev. Lett. {\bf 113}, 172301 (2014); 
{\it ibid.} Phys. Rev. D 
{\bf 91}, 045025 (2015); 
  P.~Gubler, K.~Hattori, S.~H.~Lee, M.~Oka, S.~Ozaki and K.~Suzuki,
Phys.Rev. D 93 (2016) 054026. arXiv:1512.08864 [hep-ph].
 

\bibitem{Hidaka:2012mz}
  Y.~Hidaka and A.~Yamamoto,
  Phys.\ Rev.\ D {\bf 87} (2013) 9,  094502
  [arXiv:1209.0007 [hep-ph]].
\bibitem{Luschevskaya:2014lga}
  E.~V.~Luschevskaya, O.~E.~Solovjeva, O.~A.~Kochetkov and O.~V.~Teryaev,
  Nucl.\ Phys.\ B {\bf 898} (2015) 627
  [arXiv:1411.4284 [hep-lat]].

\bibitem{Endrodi:2013cs}
  G.~Endr${\rm \ddot{o} }$di,
  JHEP {\bf 1304} (2013) 023
  [arXiv:1301.1307 [hep-ph]].

\bibitem{Agasian:2008tb}
  N.~O.~Agasian and S.~M.~Fedorov,
  Phys.\ Lett.\ B {\bf 663} (2008) 445
  doi:10.1016/j.physletb.2008.04.050
  [arXiv:0803.3156 [hep-ph]].

\bibitem{Orlovsky:2013aya}
  V.~D.~Orlovsky and Y.~A.~Simonov,
  Phys.\ Rev.\ D {\bf 89} (2014) 5,  054012
  doi:10.1103/PhysRevD.89.054012
  [arXiv:1311.1087 [hep-ph]].


\bibitem{Kojo:2013uua}
  T.~Kojo and N.~Su,
  Phys.\ Lett.\ B {\bf 726} (2013) 839
  [arXiv:1305.4510 [hep-ph]].


\bibitem{Bali:2014kia}
  G.~S.~Bali, F.~Bruckmann, G.~Endrödi, S.~D.~Katz and A.~Schäfer,
  JHEP {\bf 1408} (2014) 177
  [arXiv:1406.0269 [hep-lat]].
\bibitem{Ferrer:2012wa}
  E.~J.~Ferrer and V.~de la Incera,
  Lect.\ Notes Phys.\  {\bf 871} (2013) 399
  [arXiv:1208.5179 [nucl-th]];
  E.~J.~Ferrer, V.~de la Incera, J.~P.~Keith, I.~Portillo and P.~L.~Springsteen,
  Phys.\ Rev.\ C {\bf 82} (2010) 065802
  [arXiv:1009.3521 [hep-ph]].


\bibitem{chiralspirals}
  T.~Kojo, Y.~Hidaka, L.~McLerran and R.~D.~Pisarski,
  Nucl.\ Phys.\ A {\bf 843} (2010) 37
  [arXiv:0912.3800 [hep-ph]];
    T.~Kojo, R.~D.~Pisarski and A.~M.~Tsvelik,
  Phys.\ Rev.\ D {\bf 82} (2010) 074015
  [arXiv:1007.0248 [hep-ph]];
    T.~Kojo, Y.~Hidaka, K.~Fukushima, L.~D.~McLerran and R.~D.~Pisarski,
  Nucl.\ Phys.\ A {\bf 875} (2012) 94
  [arXiv:1107.2124 [hep-ph]].


\bibitem{Ferrer:2012zq}
  E.~J.~Ferrer, V.~de la Incera and A.~Sanchez,
  Acta Phys.\ Polon.\ Supp.\  {\bf 5} (2012) 679
  [arXiv:1205.4492 [nucl-th]].



\bibitem{EoS}
  D.~Zwanziger,
  Phys.\ Rev.\ Lett.\  {\bf 94} (2005) 182301
  [hep-ph/0407103];
   D.~Zwanziger,
  Phys.\ Rev.\ D {\bf 76} (2007) 125014
  [hep-ph/0610021];
  K.~Fukushima and N.~Su,
  Phys.\ Rev.\ D {\bf 88} (2013) 076008
  [arXiv:1304.8004 [hep-ph]];
  F.~E.~Canfora, D.~Dudal, I.~F.~Justo, P.~Pais, L.~Rosa and D.~Vercauteren,
  Eur.\ Phys.\ J.\ C {\bf 75} (2015) 7,  326
  [arXiv:1505.02287 [hep-th]].
  
\bibitem{Su:2014rma}
  N.~Su and K.~Tywoniuk,
  Phys.\ Rev.\ Lett.\  {\bf 114} (2015) 16,  161601
  [arXiv:1409.3203 [hep-ph]].
  
\bibitem{pheno}
  M.~Haas, L.~Fister and J.~M.~Pawlowski,
  Phys.\ Rev.\ D {\bf 90} (2014) 091501
  [arXiv:1308.4960 [hep-ph]];
  N.~Christiansen, M.~Haas, J.~M.~Pawlowski and N.~Strodthoff,
  Phys.\ Rev.\ Lett.\  {\bf 115} (2015) 11,  112002
  [arXiv:1411.7986 [hep-ph]];
  W.~Florkowski, R.~Ryblewski, N.~Su and K.~Tywoniuk,
  arXiv:1504.03176 [hep-ph];
  W.~Florkowski, R.~Ryblewski, N.~Su and K.~Tywoniuk,
  arXiv:1509.01242 [hep-ph];
  A.~Bandyopadhyay, N.~Haque, M.~G.~Mustafa and M.~Strickland,
  arXiv:1508.06249 [hep-ph].

\bibitem{Basar:2010zd}
  G.~Basar, G.~V.~Dunne and D.~E.~Kharzeev,
  Phys.\ Rev.\ Lett.\  {\bf 104} (2010) 232301
  [arXiv:1003.3464 [hep-ph]].


\end{thebibliography}
\end{document}